\documentstyle[onecolumn,epsfig]{mn2e}
\newif\ifAMStwofonts

\newcommand{\ica}{\sc{FastICA}}
\newcommand{\lsim}{\,\lower2truept\hbox{${<\atop\hbox{\raise4truept\hbox{$\sim$}}}$}\,}
\newcommand{\gsim}{\,\lower2truept\hbox{${>\atop\hbox{\raise4truept\hbox{$\sim$}}}$}\,}

\title[Extracting cosmic microwave background polarization from
satellite astrophysical maps] {Extracting cosmic microwave
background polarization from satellite astrophysical maps}

\author[Baccigalupi et al.]
{C. Baccigalupi$^{1,2,3}$, F. Perrotta$^{1,2,3}$, G. De
Zotti$^{4,1}$, G.F. Smoot$^{3}$, C. Burigana$^{5}$, \and
D. Maino$^{6,7}$, L. Bedini$^{8}$, E. Salerno$^{8}$\\
$^{1}$ SISSA/ISAS, Astrophysics Sector, Via Beirut, 4,
I-34014 Trieste, Italy\\
$^{2}$ INFN, Sezione di Trieste, Via Valerio 2,
I-34014 Trieste, Italy\\
$^{3}$ Lawrence Berkeley National Laboratory, 1 Cyclotron Road, Berkeley,
CA 94720, USA\\
$^{4}$ INAF, Osservatorio Astronomico di Padova,
Vicolo dell' Osservatorio 5, I-35122 Padova, Italy\\
$^{5}$ ITeSRE-CNR, Via Gobetti, 101, I-40129 Bologna, Italy\\
$^{6}$ INAF, Osservatorio Astronomico di Trieste,
Via G.B. Tiepolo, 11, I-34131 Trieste \\
$^{7}$ Dipartimento di Fisica, Universit\'a di Milano,
Via Celoria 16, I-20133, Italy \\
$^{8}$ IEI-CNR, Via Moruzzi 1, I-56124 Pisa, Italy\\
}

\pagerange{\pageref{firstpage}--\pageref{lastpage}}
\pubyear{2002}

\begin{document}

\maketitle

\label{firstpage}
\footnotetext{E-mail: bacci@sissa.it}
\begin{abstract}
We present the application of the Fast Independent Component
Analysis ({\ica}) technique for blind component separation to
polarized astrophysical emission. We study how the Cosmic
Microwave Background (CMB) polarized signal, consisting of $E$ and
$B$ modes, can be extracted from maps affected by substantial
contamination from diffuse Galactic foreground emission and instrumental 
noise. {We implement Monte Carlo chains varying the CMB and noise 
realizations in order to asses the average capabilities of the algorithm 
and their variance.} We perform the analysis of all sky maps simulated 
according to the {\sc Planck} satellite capabilities, modelling the sky
signal as a superposition of the CMB and of the existing simulated
polarization templates of Galactic synchrotron. Our results
indicate that the angular power spectrum of CMB $E$-mode can be
recovered on all scales up to $\ell\simeq 1000$, corresponding to
the fourth acoustic oscillation, while the $B$-mode power spectrum
can be detected, up to its turnover at $\ell\simeq 100$, if the
ratio of tensor to scalar contributions to the temperature
quadrupole exceeds $30\%$. The power spectrum of the cross
correlation between total intensity and polarization, $TE$, can be
recovered up to $\ell\simeq 1200$, corresponding to the seventh
$TE$ acoustic oscillation.
\end{abstract}

\begin{keywords}
methods -- data analysis -- techniques: image processing -- cosmic
microwave background.
\end{keywords}

\section{Introduction}
\label{introduction}

We are right now in the epoch in which the cosmological
observations are revealing the finest structures in the Cosmic
Microwave Background (CMB) anisotropies\footnote{see
lambda.gsfc.nasa.gov/ for the list and details of the operating
and planned CMB experiments}. After the first discovery of CMB
total intensity fluctuations as measured by the COsmic Background
Explorer (COBE) satellite (see Smoot 1999 and references therein),
several balloon-borne and ground-based operating experiments were
successful in detecting CMB anisotropies on degree and sub-degree
angular scales (De Bernardis et al 2002, Halverson et al. 2002,
Lee et al. 2001, Padin et al. 2001, see also Hu \& Dodelson 2002
and references therein). The Wilkinson Microwave Anisotropy Probe
(WMAP, see Bennett et al. 2003a)
satellite\footnote{map.gsfc.nasa.gov/} released the first year,
all sky CMB observations mapping anisotropies down to an angular
scale of about $16'$ in total intensity and its correlation with
polarization, on five frequency channels extending from 22 to 90
GHz. In the future, balloon-borne and ground based observations
will attempt to measure the CMB polarization on sky patches (see
Kovac et al. 2002 for a first detection); the {\sc
Planck}\footnote{astro.estec.esa.nl/SA-general/Projects/Planck}
satellite, scheduled for launch in 2007 (Mandolesi et al. 1998,
Puget et al. 1998), will provide total intensity and polarization
full sky maps of CMB anisotropy with resolution $\gsim 5'$ and a
sensitivity of a few $\mu$K, on nine frequencies in the 
range 30-857 GHz. A future satellite mission for polarization is
currently under study\footnote{CMBpol, see
spacescience.nasa.gov/missions/concepts.htm}.

Correspondingly, the data analysis science faces entirely new and
challenging issues in order to handle the amount of incoming data,
with the aim of extracting all the relevant physical information
about the cosmological signal and the other astrophysical
emissions, coming from extra-galactic sources as well as from our
own Galaxy. The sum of these foreground emissions, in total
intensity, is minimum at about 70 GHz, according to the first year
WMAP data (Bennett et al. 2003b). In the following we refer to low
and high frequencies meaning the ranges below and above that of
minimum foreground emission.

At low frequencies the main Galactic foregrounds are synchrotron
(see Haslam et al. 1982 for an all sky template at 408 MHz) and
free-free (traced by H$\alpha$ emission, see Haffner, Reynolds, \&
Tufte 1999, Finkbeiner 2003 and references therein) emissions, as
confirmed by the WMAP observations (Bennett et al. 2003b). At high
frequencies, Galactic emission is expected to be dominated by
thermal dust (Schlegel, Finkbeiner, Davies 1998, Finkbeiner,
Schlegel, Davies 1999). Moreover, several populations of
extra-galactic sources, with different spectral behavior, show up
at all the frequencies, including radio sources and dusty galaxies
(see Toffolatti et al. 1998), and the Sunyaev-Zeldovich effect
from clusters of galaxies (Moscardini et al. 2002). Since the
various emission mechanisms have generally different frequency
dependencies, it is conceivable to combine multi-frequency maps in
order to separate them.

A lot of work has been recently dedicated to provide algorithms
devoted to the component separation task, exploiting different
ideas and tools from signal processing science. Such algorithms
generally deal separately with point-like objects like
extra-galactic sources (Tenorio et al. 1999, Vielva et al. 2001),
and diffuse emissions from our own Galaxy. In this work we focus
on techniques developed to handle diffuse emissions; such
techniques can be broadly classified in two main categories.

The ``non-blind" approach consists in assuming priors on the
signals to recover, on their spatial pattern and frequency
scalings, in order to regularize the inverse filtering going from
the noisy, multi-frequency data to the separated components.
Wiener filtering (WF, Tegmark, Efstathiou 1996, Bouchet, Prunet,
Sethi 1999) and Maximum Entropy Method (MEM, Hobson et al. 1998)
have been tested with good results, even if applied to the whole
sky (Stolyarov et al. 2002). Part of the priors can be obtained
from complementary observations, and the remaining ones have to be
guessed. The WMAP group (Bennett et al. 2003b) exploited the
available templates mentioned above as priors for a successful
MEM-based component separation.

The ``blind" approach consists instead in performing separation by
only assuming the statistical independence of the signals to
recover, without priors either for their frequency scalings, or
for their spatial statistics. This is possible by means of a novel
technique in signal processing science, the Independent Component
Analysis (ICA, see Amari \& Chichocki 1998 and references
therein). The first astrophysical application of this technique
(Baccigalupi et al. 2000) exploited an adaptive (i.e. capable of
self-adjusting on time streams with varying signals) ICA
algorithm, working successfully on limited sky patches for ideal
noiseless data. Maino et al. (2002) implemented a fast,
non-adaptive version of such algorithm ({\ica}, see Hyv\"{a}rinen
1999) which was successful in reaching separation of CMB and
foregrounds for several combinations of simulated all sky maps in
conditions corresponding to the nominal performances of {\sc
Planck}, for total intensity measurements. Recently, Maino et al.
(2003) were able to reproduce the main scientific results out of
the COBE data exploiting the {\ica} technique. The blind
techniques for component separation represent the most unbiased
approach, since they only assume the statistical independence
between cosmological and foreground emissions. Thus they not only
provide an independent check on the results of non-blind
separation procedures, but are likely to be the only viable way to
go when the foreground contamination is poorly known.

In this work we apply the {\ica} technique to astrophysical
polarized emission. CMB polarization is expected to arise from
Thomson scattering of photons and electrons at decoupling. Due to
the tensor nature of polarization, physical information is coded
in a entirely different way with respect to total intensity.
Cosmological perturbations may be divided into scalars, like
density perturbations, vectors, for example vorticity, and
tensors, i.e. gravitational waves (see Kodama \& Sasaki 1984).
Total intensity CMB anisotropies simply sum up contributions from
all kinds of cosmological perturbations. For polarization, two
non-local combinations of the Stokes parameters $Q$ and $U$ can be
built, commonly known as $E$ and $B$ modes (see Zaldarriaga, \&
Seljak 1997, and Kamionkowski, Kosowsky \& Stebbins 1997 featuring
a different notation, namely gradient $G$ for $E$ and curl $C$ for
$B$). It can be shown that the $E$ component sums up the
contributions from all the three kinds of cosmological
perturbations mentioned above, while the $B$ modes are excited via
vectors and tensors only. Also, scalar modes of total intensity,
which we label with $T$ in the following, are expected to be
strongly correlated with $E$ modes: indeed, the latter are merely
excited by the quadrupole of density perturbations, coded in the
total intensity of CMB photons, as seen from the rest frame of
charged particles at last scattering (see Hu et al. 1999 and
references therein). Therefore, for CMB, the correlation $TE$
between $T$ and $E$ modes is expected to be the strongest signal
from polarization. The latter expectation has been confirmed by
WMAP (Kogut et al. 2003) with a spectacular detection on degree and 
super-degree angular scales; moreover, a first detection of CMB $E$
modes has been obtained (Kovac et al. 2002).

This phenomenology is clearly much richer with respect to total
intensity, and motivated a great interest toward CMB polarization,
not only as a new data set in addition to total intensity, but as
the best potential carrier of cosmological information via
electromagnetic waves. Unfortunately, as we describe in the next
Section, foregrounds are even less known in polarization than in
total intensity, see De Zotti (2002) and references therein for
reviews. For this reason, it is likely that a blind technique will
be required to clean CMB polarization from contaminating
foregrounds. The first goal of this work is to present a first
implementation of the ICA techniques on polarized astrophysical
maps. Second, we want to estimate the precision with which CMB
polarized emission will be measured in the near future. We exploit
the {\ica} technique on low frequencies where some foreground
model have been carried out (Giardino et al. 2002, Baccigalupi et
al. 2001).

The paper is organized as follows. In Section \ref{simulated} we
describe how the simulation of the synchrotron emission were
obtained. In Section \ref{component} we describe our approach to
component separation for polarized radiation. In Section
\ref{performance} we study the {\ica} performance on our simulated
sky maps. In Section \ref{application} we apply our technique to
the {\sc Planck} simulated data, studying its capabilities for
polarization measurements in presence of foreground emission.
Finally, Section \ref{concluding} contains the concluding remarks.

\section{Simulated polarization maps at microwave frequencies}
\label{simulated}

We adopt a background cosmology close to the model which best fits
the WMAP data (Spergel et al. 2003). We assume a flat Friedman
Robertson Walker (FRW) metric with an Hubble constant $H_{0}=100h$
km/sec/Mpc with $h=0.7$. The Cosmological Constant represents
$70\%$ of the critical density today, $\Omega_{\Lambda}=0.7$,
while the energy density in baryons is given by
$\Omega_{b}h^{2}=0.022$; the remaining fraction is in Cold Dark
Matter (CDM); we allow for a re-ionization with optical depth
$\tau =0.05$ (Becker et al. 2001). Note that this is a factor 2-3
lower than found in the first year WMAP data (see Bennett et al.
2003a and references therein), since we built our reference CMB 
template before the release of WMAP data. Cosmological
perturbations are Gaussian, with spectral index for the scalar
component leading to a not perfectly scale invariant spectrum,
$n_{S}=0.96$, and including tensor perturbations giving rise to a
$B$ mode in the CMB power spectrum. We assume a ratio $R=30\%$
between tensor and scalar amplitudes, and the tensor spectral
index is taken to be $n_{T}=-R/6.8$ according to the simplest
inflationary models of the very early Universe (see Liddle \& Lyth
2000 and references therein). The cosmological parameters leading
to our CMB template can be summarized as follows:
\begin{equation}
\label{CMB}
h=0.7\ .\ \Omega_{\Lambda}=0.7\ ,\ \Omega_{b}h^{2}=0.022
\ ,\ \Omega_{CDM}=1-\Omega_{\Lambda}-\Omega_{b}\ ,\ \tau =0.05
\ ,\ n_{S}=0.96\ ,\ R=0.3\ ,\ n_{T}=-R/6.8\ .
\end{equation}
We simulate whole sky maps of $Q$ and $U$ out of the theoretical
$C_{\ell}^{E}$ and $C_{\ell}^{B}$ coefficients as generated by
{\sc CMBfast} (Seljak \& Zaldarriaga 1996), in the HEALPix
environment (G\'orski et al. 1999). The maps are in antenna
temperature, which is obtained at any frequency $\nu$ multiplying
the thermodynamical fluctuations by a factor
$x^{2}\exp{x}/(\exp{x}-1)^{2}$, where $x=h\nu /kT_{CMB}$, $h,k$
are the Planck and the Boltzmann constant, respectively, while
$T_{CMB}=2.726$ K is the CMB thermodynamical temperature.

The polarized emission from diffuse Galactic foregrounds in the
frequency range which will be covered by the {\sc Planck}
satellite is very poorly known. On the high frequency side the
Galactic contribution to the polarized signal should be dominated
by dust emission (Lazarian \& Prunet 2002). The first detection of
the diffuse polarized dust emission has been carried out recently
(Benoit et al. 2003), and indicates a $3-5\%$ polarization on
large angular scales and at low Galactic latitudes. On the low
frequency side, the dominant diffuse polarized emission is
Galactic synchrotron. Observations in the radio band cover about
half of the sky at degree resolution (Brouw \& Spoelstra 1976),
and limited regions at low and medium Galactic latitudes with
$10'$ resolution (Duncan et al. 1997, Uyaniker et al. 1999, Duncan
et al. 1999). Analyses of the angular power spectrum of polarized
synchrotron emission have been carried out by several authors
(Tucci et al. 2000, Baccigalupi et al. 2001, Giardino et al. 2002,
Tucci et al. 2002, Bernardi et al. 2003).

Polarized foreground contamination is particularly challenging for
CMB $B$-mode measurements. In fact, the CMB $B$ mode arises from
tensor perturbations (see Liddle \& Lyth 2000 for reviews), which
are subdominant with respect to the scalar component (Spergel et
al. 2003). In addition, tensor perturbations vanish on sub-degree
angular scales, corresponding to sub-horizon scales at decoupling.
On such scales, some $B$-mode power could be introduced by weak
lensing (see Hu 2002 and references therein). Anyway, the
cosmological $B$-mode power is always expected to be much lower
than  the $E$-mode, while foregrounds are expected to have
approximately the same power in the two modes (Zaldarriaga 2001).

Baccigalupi et al. (2001) estimated the power spectrum of
synchrotron as derived by two main data-sets. As we already
mentioned, on super-degree angular scales, corresponding to
multipoles $\ell < 200$, the foreground contamination is
determined from the Brouw \& Spoelstra (1976) data, covering
roughly half of the sky with degree resolution. The $C_{\ell}$
behavior on smaller angular scales has been obtained analyzing
more recent data reaching a resolution of about $10'$ (Duncan et
al. 1997, 1999, Uyaniker et al. 1999). These data, reaching
Galactic latitudes up to $b\simeq 20^{\circ}$, yield a flatter
slope, $C_{\ell}\simeq \ell^{-(1.5 \div 1.8)}$ (see also Tucci et
al. 2000, Giardino et al. 2002) . Fosalba et al. (2002)
interestingly provided evidence of a similar slope for the angular
power spectrum of the polarization degree induced by the Galactic
magnetic field as measured from starlight data. The synchrotron
spectrum at higher frequencies was then inferred
by scaling the one obtained in the radio band with a typical
synchrotron spectral index of $-2.9$ (in antenna temperature).

\begin{figure}
\centerline{
\epsfig{file=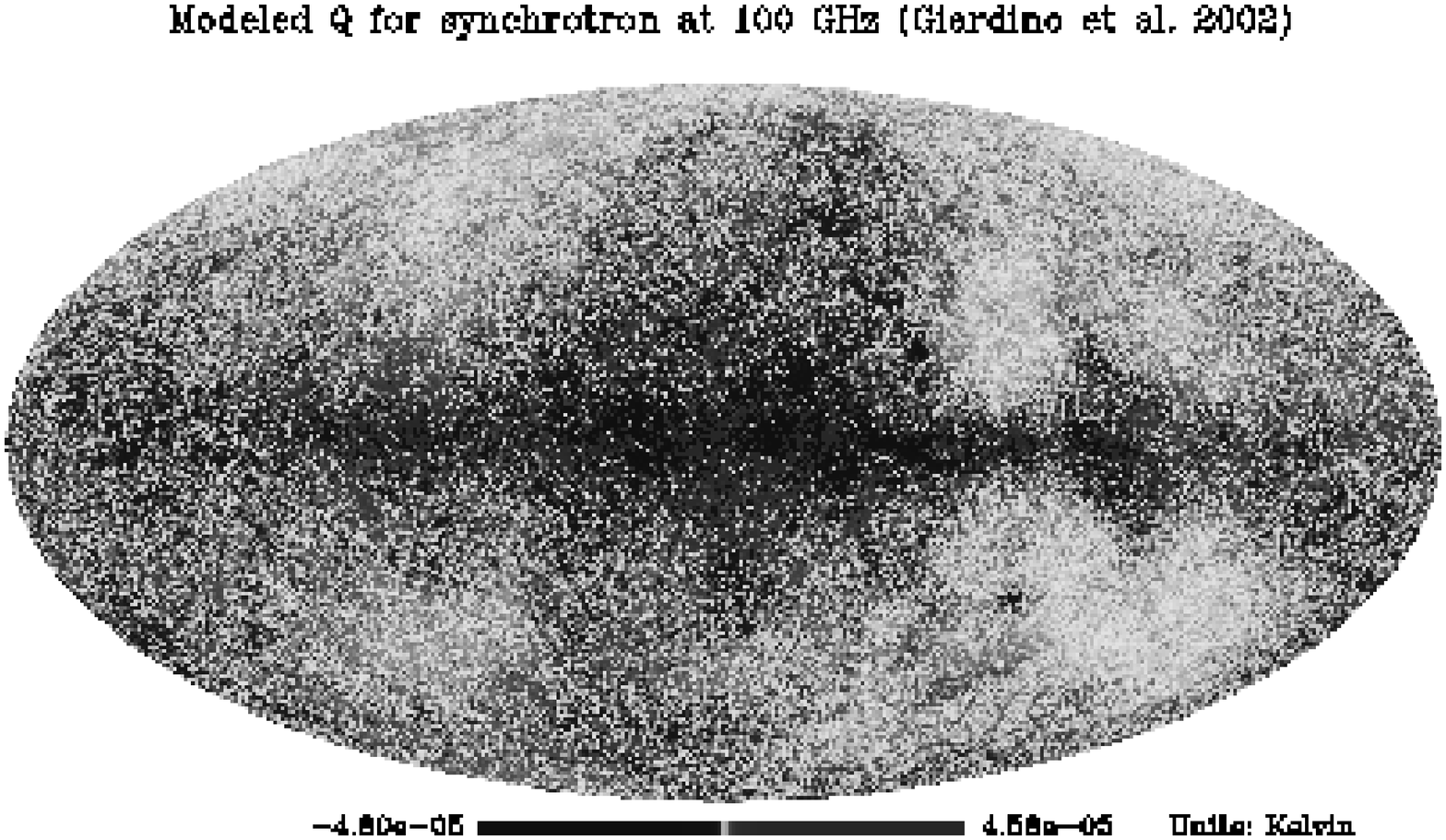,height=3.in,width=2.in,angle=90}
\epsfig{file=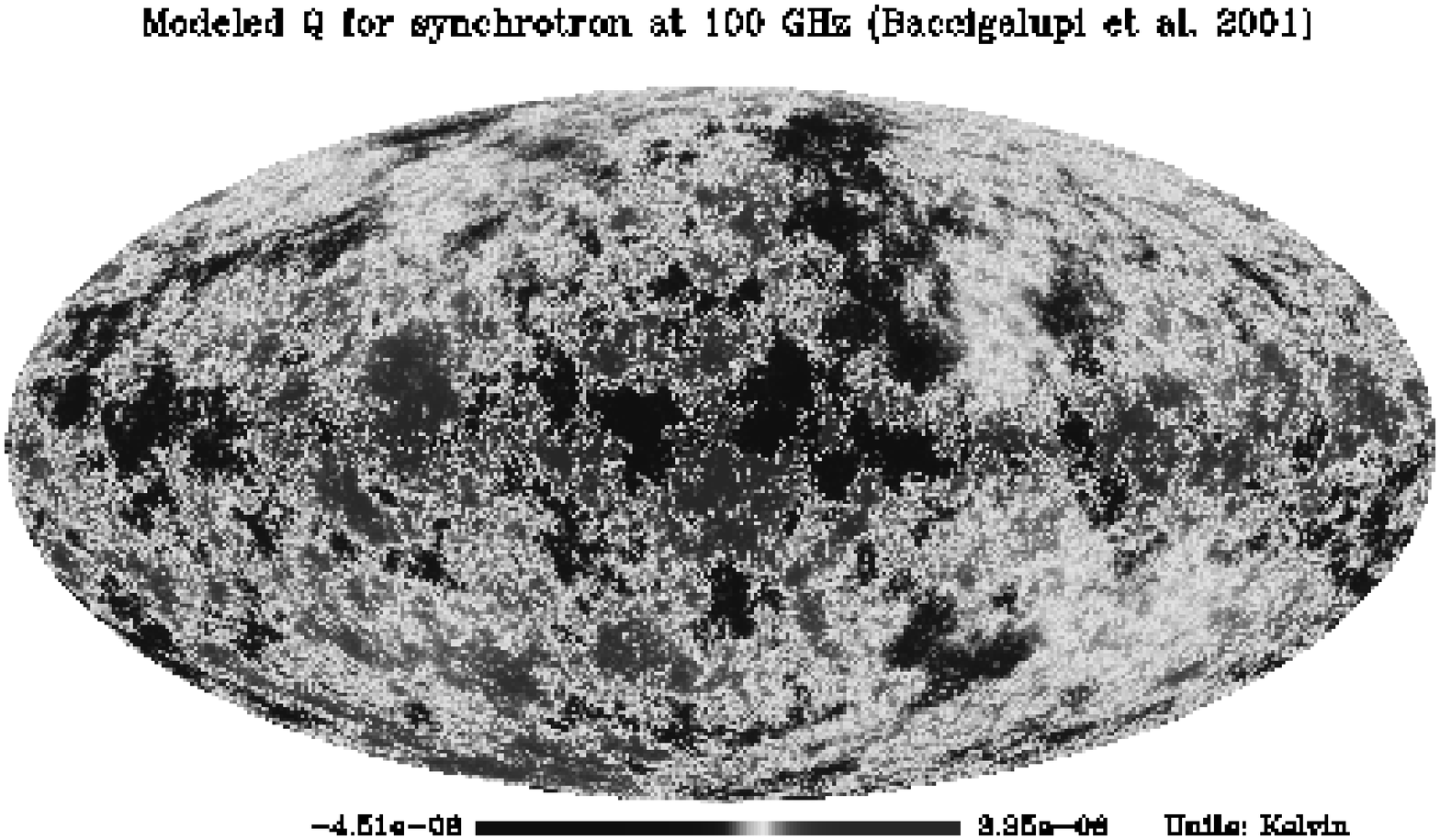,height=3.in,width=2.in,angle=90}}
\caption{Q Stokes parameter for the emission of Galactic
synchrotron according to Giardino et al. (2002, left panel)
and Baccigalupi et al. (2001, right panel). The maps are
in antenna temperature, at 100 GHz.}
\label{synsky}
\end{figure}

Giardino et al. (2002) built a full sky map of synchrotron
polarized emission based on the total intensity map by Haslam et
al. (1982), assuming a synchrotron polarized component at the
theoretical maximum level of $75\%$, and a Gaussian distribution
of polarization angles with a power spectrum estimated out of the
high resolution radio band data (Duncan et al. 1997, 1999; Duncan
et al. 1999). The polarization map obtained, reaching a resolution
of about $10'$, was then scaled to higher frequencies by
considering either a constant or a space-varying spectral index as
inferred by multi-frequency radio observations.

\begin{figure}
\centerline{
\epsfig{file=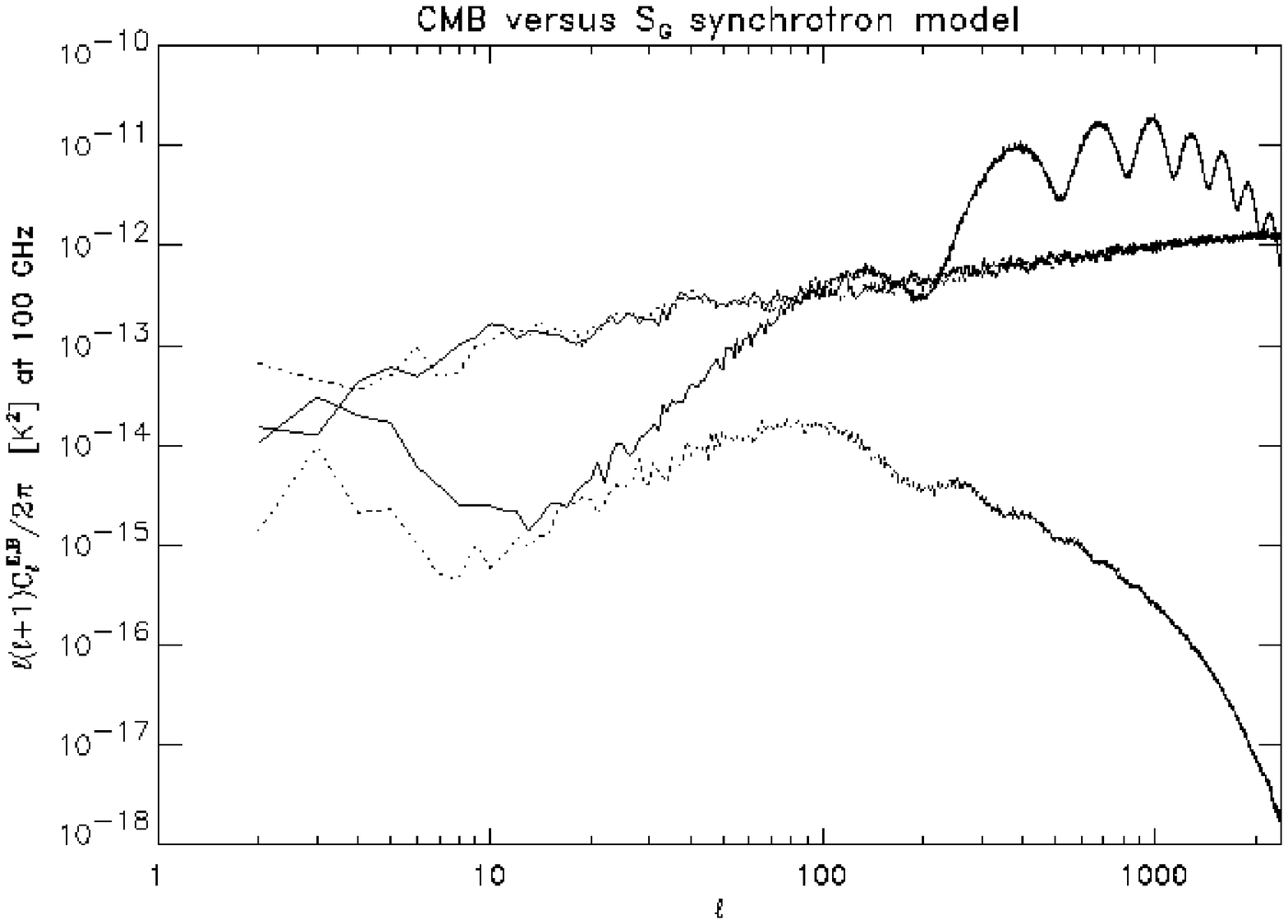,height=3.in,width=3.in}
\epsfig{file=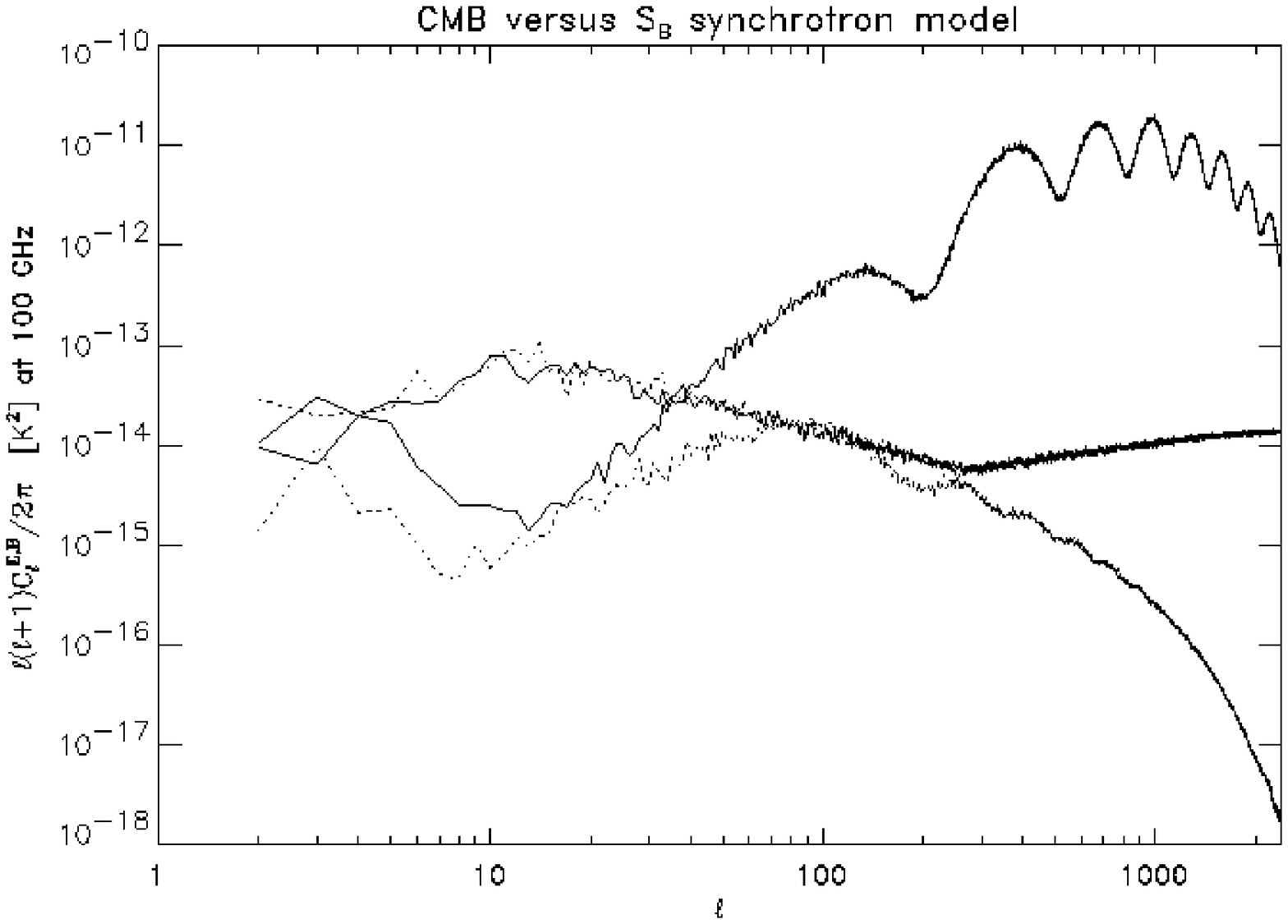,height=3.in,width=3.in} }
\caption{$E$ (solid) and $B$ (dotted) angular power spectra of CMB
and synchrotron polarization emission, in antenna temperature, at
100 GHz, according to the $S_{G}$ (left panel), and $S_{B}$ (right
panel) polarized synchrotron template.} \label{clsynsky}
\end{figure}

In this work we concentrate on low frequencies, modelling the
diffuse polarized emission as a superposition of CMB and
synchrotron. We used the synchrotron spatial template by Giardino
et al. (2002), hereafter $S_{G}$ model, as well as another
synchrotron template, hereafter indicated as $S_{B}$, obtained by
scaling the spherical harmonics coefficients of the $S_{G}$ model
to match the spectrum found by Baccigalupi et al. (2001). The $Q$
Stokes parameter for the two spatial templates, in antenna
temperature at 100 GHz, are shown in Fig. \ref{synsky}, plotted in
a non-linear scale to highlight the behavior at high Galactic
latitudes. Note how the contribution on smaller angular scales is
larger in the $S_{G}$ model. This is evident in
Fig.~\ref{clsynsky} where we compare the power spectra of the
$S_{G}$ and $S_{B}$ models with the CMB one, for the cosmological
parameters of Eq.~(\ref{CMB}). Both models imply a severe
contamination of the CMB $E$ mode on large angular scales, say
$\ell\lsim 200$, which remains serious even if the Galactic plane
is cut out: cutting the region $|b|\le 20^{\circ}$ decreases the
$S_{G}$ and $S_{B}$ signals by about a factor of 10 and 3,
respectively; the difference comes from the fact that the $S_{B}$
power is concentrated more on large angular scales, which
propagate well beyond the Galactic plane (see
Fig.~\ref{clsynsky}).

In the $S_{G}$ case the contamination is severe also for the first
CMB acoustic oscillation in polarization, as a result of the
enhanced power on small angular scales with respect to the $S_{B}$
models (see also Fig.~\ref{synsky}). On smaller scales both models
predict the dominance of CMB $E$ modes. On the other hand, CMB $B$
modes are dominated by foreground emission, even if the region
around the Galactic plane is cut out as we commented above.

\begin{figure}
\centerline{
\epsfig{file=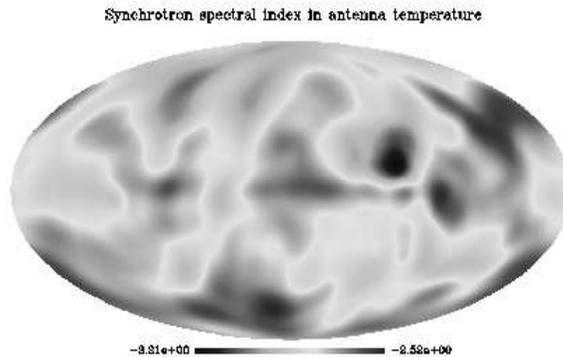,height=3.in,width=2.in,angle=90} }
\caption{Map of synchrotron spectral indices (Giardino et al.
2002).} \label{synspecind}
\end{figure}

To obtain the synchrotron emission at different frequencies, we
consider either a constant antenna temperature spectral index of
$-2.9$, slightly shallower than indicated by the WMAP first year
measurements (Bennett et al. 2003b), as well as a varying spectral
index. In Fig.~\ref{synspecind} we show the map of synchrotron
spectral indices, in antenna temperature, which we adopt following
Giardino et al. (2002). Note that this aspect is relevant
especially for component separation, since all methods developed
so far require a ``rigid" frequency scaling of all the components,
which means that all components should have separable dependencies
on sky direction and frequency. Actually this requirement is hardly 
satisfied by real signals, and by synchrotron in particular.
However, as we see in the next Section, {\ica} results turn out to
be quite stable as this assumption is relaxed, at least for the
level of variation in Fig.~\ref{synspecind}. This make this
technique very promising for application to real data. A more
quantitative study of how the {\ica} performance gets degraded
when realistic signals as well instrumental systematics are taken
into account will be carried out in a future work.

\section{Component separation for polarized radiation}
\label{component}

Component separation has been implemented so far for the total
intensity signal (see Maino et al. 2002, Stolyarov et al. 2002,
and references therein). In this Section we expose how we extend
the ICA technique to treat polarization measurements.

\subsection{$E$ and $B$ modes}
\label{e}

As we stressed in the previous Section, the relevant information
for the CMB polarized signal can be conveniently read in a
non-local combination of $Q$ and $U$ Stokes parameters,
represented by the $E$ and $B$ modes (see Zaldarriaga \& Seljak
1997 and Kamionkowski, Kosowsky, \& Stebbins 1997). {There are
conceptually two ways of performing component separation in
polarization observations. $Q$ and $U$ can be treated separately,
i.e. performing separation for each of them independently.
However, in the hypothesis that $Q$ and $U$ have the same
statistical properties, separation can be conveniently performed
on a data-set combining $Q$ and $U$ maps. This is surely the
appropriate strategy if one is sure that the choice of
polarization axes of the instrumental set-up does not bias the
signal distribution. In general, however, it may happen that
accidentally the instrumental polarization axes are related to the
preferred directions of the underlying signal, making $Q$ and $U$
statistically different, so that merging them in a single data-set
would not be appropriate. While for the Gaussian CMB statistics we
do not expect such occurrence, it may happen for foregrounds, 
expecially if separation is
performed on sky patches. For example, the Galactic polarized
signal possesses indeed large scale structures with preferred
directions, such as the Galactic Plume, discovered by Duncan et
al. (1998) in the radio band, extending up to 15 degrees across
the sky and reaching high Galactic latitudes. Therefore, in
general, the most conservative approach to component separation in
polarization is to perform it for $Q$ and $U$ separately. Note
that in our Galactic model no coherence in the polarization
detection is present. This allow us to verify, in the next
Section, that the results obtained by merging $Q$ and $U$ in a
single data-set are quite equivalent or more accurate than those
obtained by treating them separately. In the following we report
the relevant {\ica} formalism for the latter case. Otherwise, when
$Q$ and $U$ form a single template, the same formulas developed in
Maino et al. (2002) do apply.}

Let these multi-frequency maps be represented by ${\bf x}^{Q}$ and
${\bf x}^{U}$ respectively, where ${\bf x}$ is made of two
indexes, labelling frequencies on rows and pixel on columns. If
the unknown components to be recovered from the input data scale
rigidly in frequency, which means that each of them can be
represented by a product of two functions depending on frequency
and space separately, we can define a spatial pattern for them,
which we indicate with ${\bf s}^{Q}$ and ${\bf s}^{U}$. Then we
can express the inputs ${\bf x}^{Q,U}$ as
\begin{equation}
\label{xQU}
{\bf x}^{Q,U}={\bf A}^{Q,U}{\bf B}\, {\bf s}^{Q,U}+{\bf n}^{Q,U}\ ,
\end{equation}
where the matrix ${\bf A}^{Q,U}$ scales the spatial patterns of
the unknown components to the input frequencies, thus having a
number of rows equal to the number of input frequencies. The
instrumental noise ${\bf n}$ has same dimensions as ${\bf x}$. The
matrix {\bf B} represents the beam smoothing operation: we recall
that at the present level of architecture, an ICA based component
separation requires to deal to maps having equal beams at all
frequencies. Separation is achieved in the real space, by
estimating two separation matrices, ${\bf W}^{Q}$ and ${\bf
W}^{U}$, having a number of rows corresponding to the number of
independent components and a number of columns equal to the
frequency channels, which produce a copy of the independent
components present in the input data:
\begin{equation}
\label{yQU}
{\bf y}^{Q}={\bf W}^{Q}{\bf x}^{Q}\ \ ,\ \ {\bf y}^{U}={\bf W}^{U}{\bf x}^{U}\ .
\end{equation}
All the details on the way the separation matrix for {\ica} is
estimated are given in Maino et al. (2002). ${\bf y}^{Q}$ and
${\bf y}^{U}$ can be combined together to get the $E$ and $B$
modes for the independent components present in the input data
(see Zaldarriaga, Seljak 1997 and Kamionkowski, Kosowsky, Stebbins
1997). Note that failures in separation for even one of $Q$ and
$U$ affect in general both $E$ and $B$, since each of them receive
contributions from both $Q$ and $U$. It is possible, even in the
noisy case, to check the quality of the resulting separation by
looking at the product ${\bf W}{\bf A}$, which should be the
identity in the best case. This means that the frequency scalings
of the recovered components can be estimated: following Maino et
al. (2002), by denoting as $x_{\nu\, j}^{Q,U}$ the $j$-th
component in the data {\bf x} at frequency $\nu$, it can be easily
seen that the frequency scalings are simply the ratios of the
column elements of the matrices {\bf W$^{-1}$}:
\begin{equation}
\label{freqscal}
{x_{\nu\, j}^{Q,U}\over x_{\nu'\, j}^{Q,U}}=
{({\rm\bf W}^{Q,U})^{-1}_{\nu\, j}\over
({\rm\bf W}^{Q,U})^{-1}_{\nu'\, j}}\ .
\end{equation}
However, it is important to note that even if separation goes
virtually perfect, which means that ${\bf W}{\bf A}$ is exactly
the identity, Eqs.~(\ref{xQU},\ref{yQU}) imply that noise is
transmitted to the {\ica} outputs, even if it can be estimated
and, to some extent, taken into account during the separation
process.

\subsection{Instrumental noise}
\label{instrumental}

Our method to deal with instrumental noise in a {\ica}-based
separation approach is described in Maino et al. (2002), for total
intensity maps. Before starting the separation process, the noise
correlation matrix, which for a Gaussian, uniformly distributed
noise is null except for the noise variances at each frequency on
the diagonal, is subtracted from the total signal correlation
matrix; the ``denoised" signal correlation matrix enters then as
an input in the algorithm performing separation. The same is done
also here, for $Q$ and $U$ separately. Moreover, in Maino et al.
(2002) we described how to estimate the noise of the {\ica}
outputs. In a similar way, let us indicate the input noise
patterns as ${\bf n}^{Q,U}$. Then from Eqs.~(\ref{xQU},\ref{yQU})
it can be easily seen that the noise on {\ica} outputs is given by
\begin{equation}
\label{nyQU}
{\bf n}_{y}^{Q,U}={\bf W}^{Q,U}{\bf n}^{Q,U}\ ,
\end{equation}
which means that, if the noises on different channels are uncorrelated,
and indicating as $\sigma_{\nu_{j}}$ the input noise $rms$ at
frequency $\nu_{j}$, the noise $rms$ on the $i$-th {\ica} output is
\begin{equation}
\label{sigmayQU}
\sigma_{y_{i}}^{Q,U}=
\sqrt{\sum_{j}|W_{ij}^{Q,U}|^{2}|\sigma_{\nu_{j}}^{Q,U}|^{2}}\ .
\end{equation}
Note that the above equation describes the amount of noise 
which is transmitted to the outputs after the separation matrix 
has been found, and not how much the separation matrix is affected 
by the noise. As treated in detail in Maino et al. (2002), if the 
noise correlation matrix is known, it is possible to subtract it 
from the signal correlation matrix, greatly reducing the influence 
of the noise on the estimation of ${\bf W}^{Q,U}$; however, sample 
variance and in general any systematics will make the separation 
matrix noisy in a way which is not accounted by Eq.(\ref{sigmayQU}). \\
On whole sky signals, the contamination to the angular power
spectrum coming from a uniformly distributed, Gaussian noise
characterized by $\sigma$ is $C_{\ell}=4\pi\sigma^{2}/N$, where
$N$ is the pixel number on the sphere. The noise contamination to
$C_{\ell}$s on $Q$ and $U$ can therefore be estimated easily on
{\ica} outputs once $\sigma_{\nu_{j}}^{Q,U}$ in
Eq.~(\ref{sigmayQU}) are known. Gaussianity and uniformity make
also very easy to calculate the noise level on $E$ and $B$ modes,
since they contribute at the same level. Thus we can estimate the
noise contamination in the $E$ and $B$ channels as
\begin{equation}
\label{clnEB}
C_{\ell\, y_{i}}^{E}=C_{\ell\, y_{i}}^{B}=
{C_{\ell\, y_{i}}^{Q}+C_{\ell\, y_{i}}^{U}\over 4}=
{\pi (|\sigma_{y_{i}}^{Q}|^{2}+|\sigma_{y_{i}}^{U}|^{2})\over N}\ ,
\end{equation}
where the factor 4 is due to the normalization according to the
HEALPix scheme (version 1.10 and less), featuring conventions of 
Kamionkowski, Kosowsky,
\& Stebbins (1997), whereas the other common version (Zaldarriaga
\& Seljak 1997) would yield a factor 2. The quantities defined in
Eq.~(\ref{clnEB}) represent the average noise power, which can be
simply subtracted from the output power spectra by virtue of the
uncorrelation between noise and signal; the noise contamination is
then represented by the power of noise fluctuations around the
mean [Eq.~(\ref{clnEB})]:
\begin{equation}
\Delta C_{\ell\, y_{i}}^{E}=\Delta C_{\ell\, y_{i}}^{B}=
\sqrt{2\over 2\ell +1}\left[
{\pi (|\sigma_{y_{i}}^{Q}|^{2}+|\sigma_{y_{i}}^{U}|^{2})\over N}\right]\ .
\label{deltaclnEB}
\end{equation}
Note that the noise estimation is greatly simplified by our
assumptions: a non-Gaussian and/or non-uniform noise, as well as a
non-zero $Q/U$ noise correlation etc. could lead for instance to
non-flat noise spectra for $C_{\ell}$s, as well as non-equal
noises in $E$ and $B$ modes. However, if a good model of these
effects is available, a Monte Carlo pipeline is still conceivable
by calculating many realizations of noise to find the average
contamination to $E$ and $B$ modes to be subtracted from outputs
instead of the simple forms of
Eqs.~(\ref{clnEB},\ref{deltaclnEB}).

\section{Performance study}
\label{performance}

In this Section we apply our approach to simulated skies to
assess: $(i)$ the ultimate capability of {\ica} to clean the CMB
maps from synchrotron in ideal noise-less conditions, $(ii)$ how
the results are degraded by noise.

\subsection{Noiseless separation}
\label{noiseless}

We work with angular resolution of $3'.5$, corresponding to
$n_{side}=1024$ in an HEALPix environment (G\'orski et al. 1999);
this is enough to test the performance of the CMB polarization
reconstruction, in particular for the undamped sub-degree acoustic
oscillations, extending up to $\ell\simeq 2000$ in
Fig.~\ref{clsynsky}. In all the cases we show, the computing time
to achieve separation was of the order of a few minutes on a
Pentium IV 1.8 GHz processor with 512 Mb RAM memory. We perform
separation by considering the CMB model defined by Eq.~(\ref{CMB})
and both the $S_{G}$ (Giardino et al. 2002) and the $S_{B}$
(Baccigalupi et al. 2001) model for synchrotron emission. {We
have considered $Q$ and $U$ maps both separately and combined.}

\begin{figure*}
\begin{center}
\epsfig{file=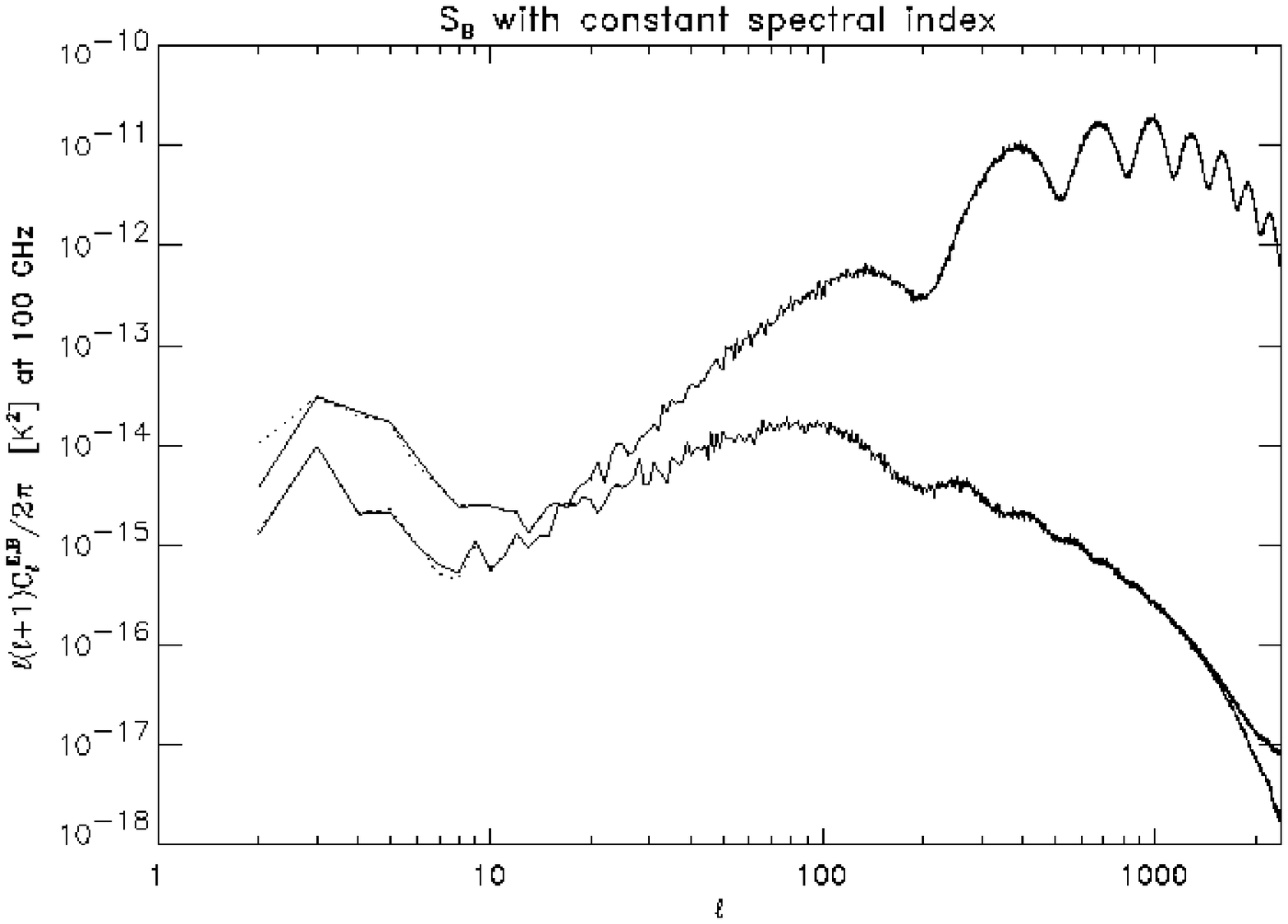,height=2.5in,width=3.in,angle=0}
\hskip .2in
\epsfig{file=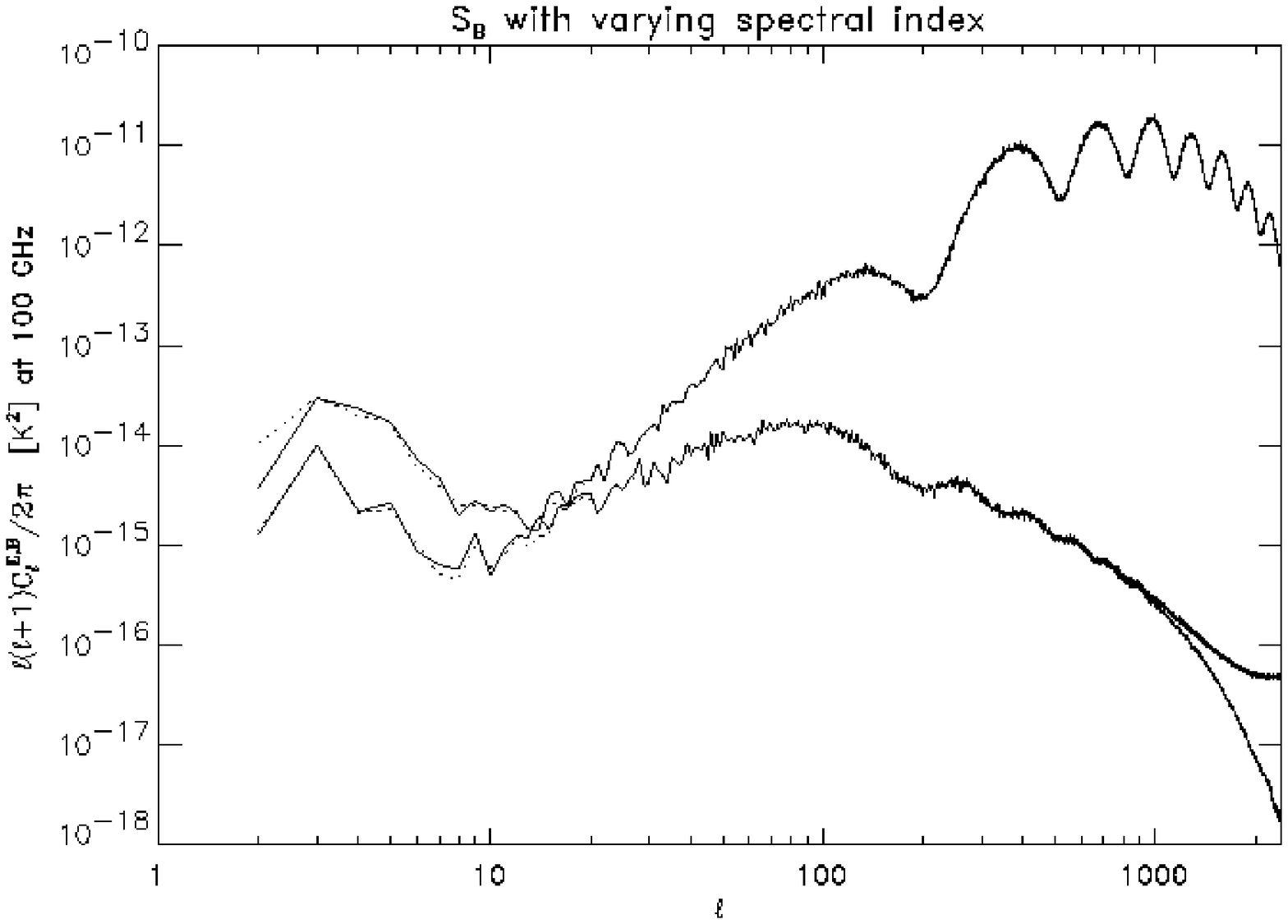,height=2.5in,width=3.in,angle=0}
\vskip .3in
\epsfig{file=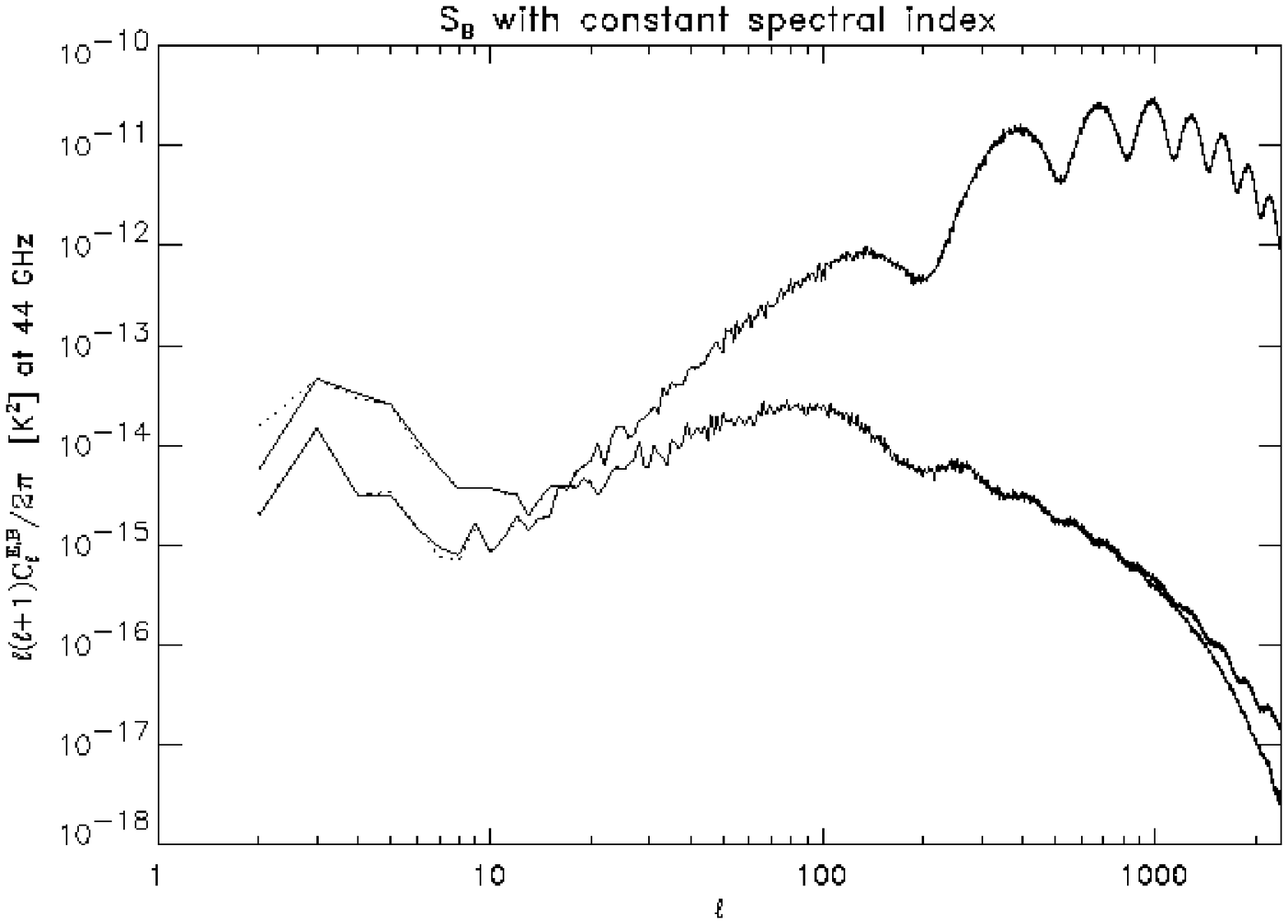,height=2.5in,width=3.in,angle=0}
\hskip .2in
\epsfig{file=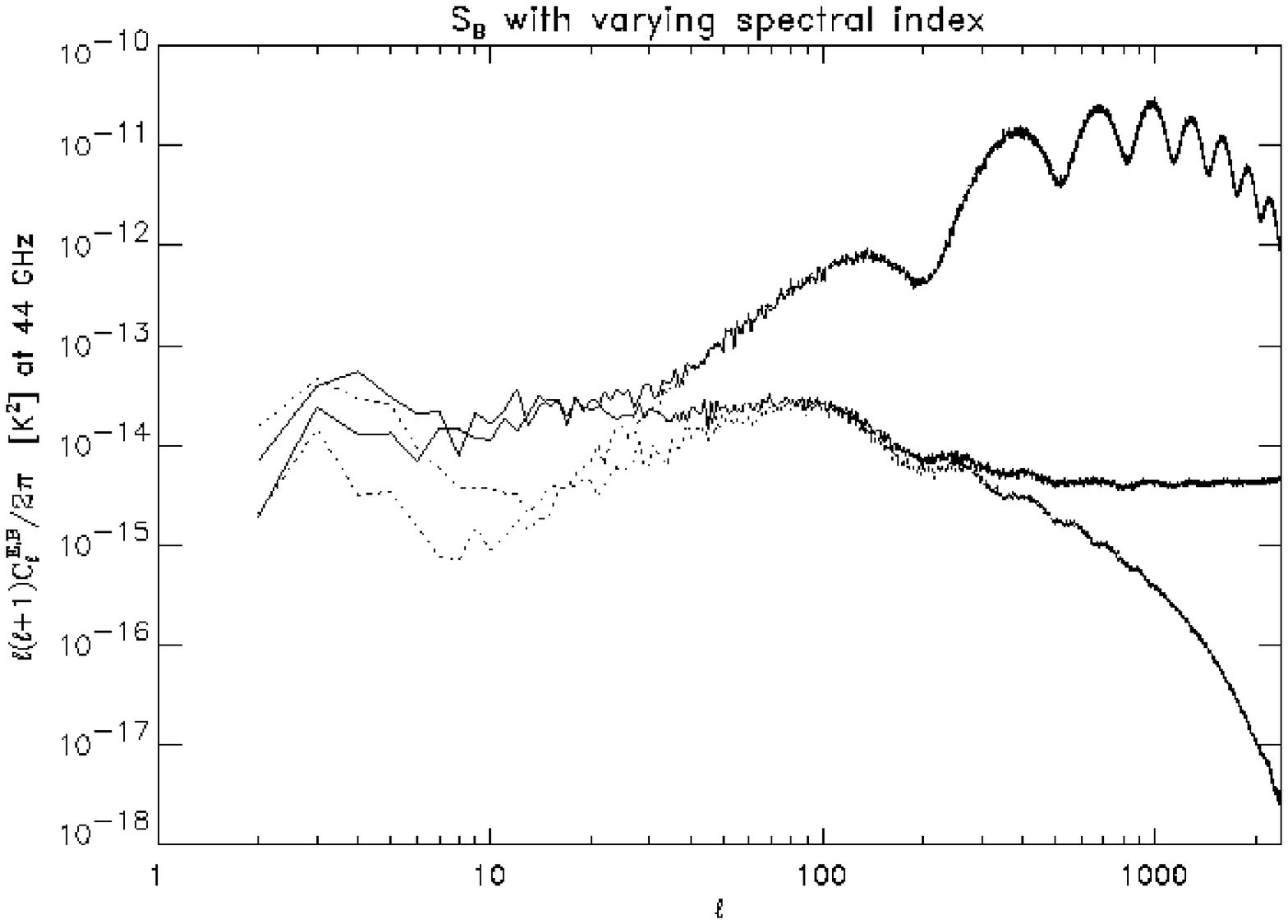,height=2.5in,width=3.in,angle=0}
\vskip .3in \caption{Original (dotted) and reconstructed (solid)
power spectra for $E$ and $B$ CMB modes in the case of the $S_{B}$
model, in the absence of noise. Left panels: constant spectral
index. Right panels: space varying spectral index. Upper panels
are from inputs at 70 and 100 GHz channels, while bottom panels
corresponds to inputs at 30 and 44 GHz. The outputs are conventionally 
plotted at the highest input frequency.}
\label{nonoisemodellobacci}
\end{center}
\end{figure*}

\begin{figure*}
\begin{center}
\epsfig{file=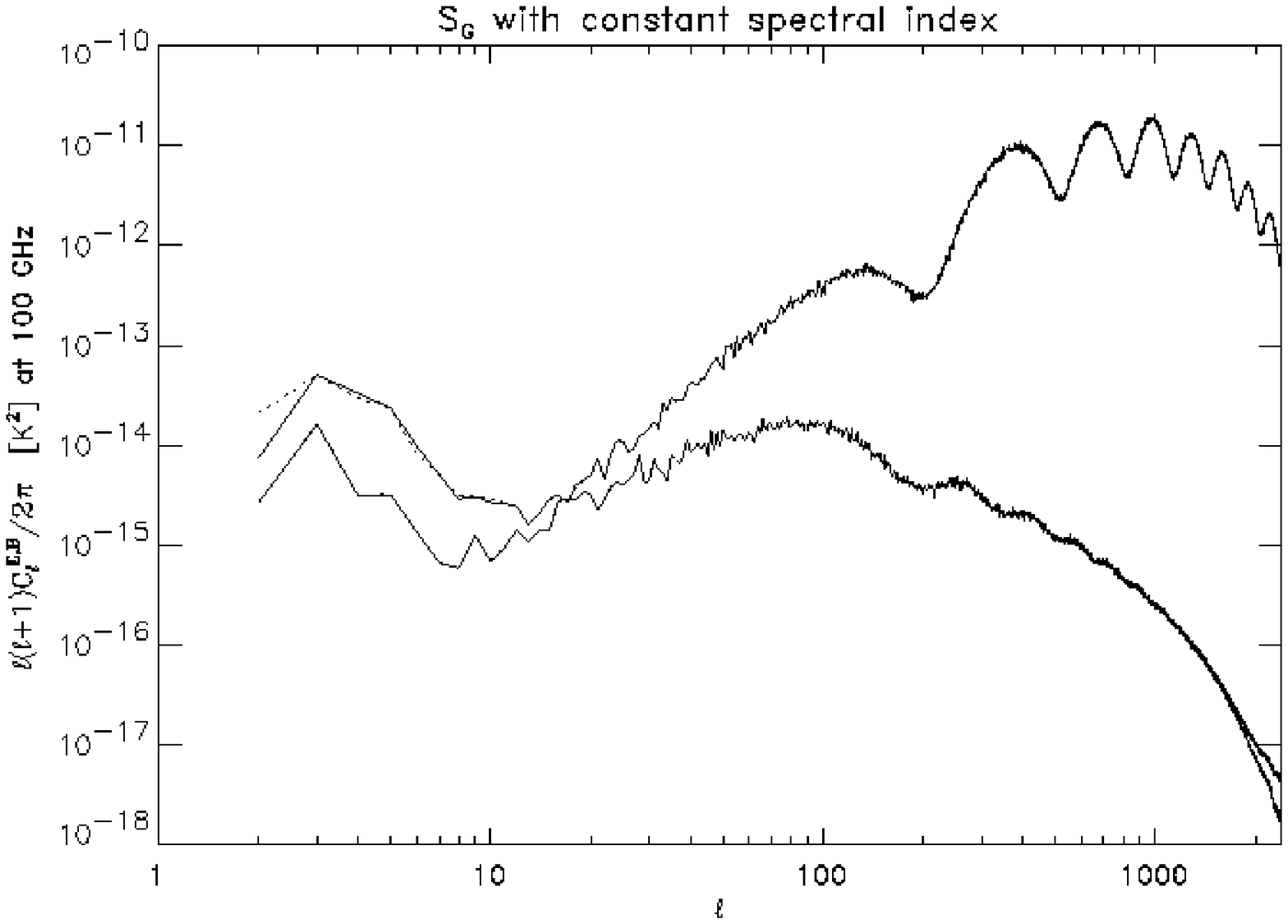,height=2.5in,width=3.in,angle=0}
\hskip .2in
\epsfig{file=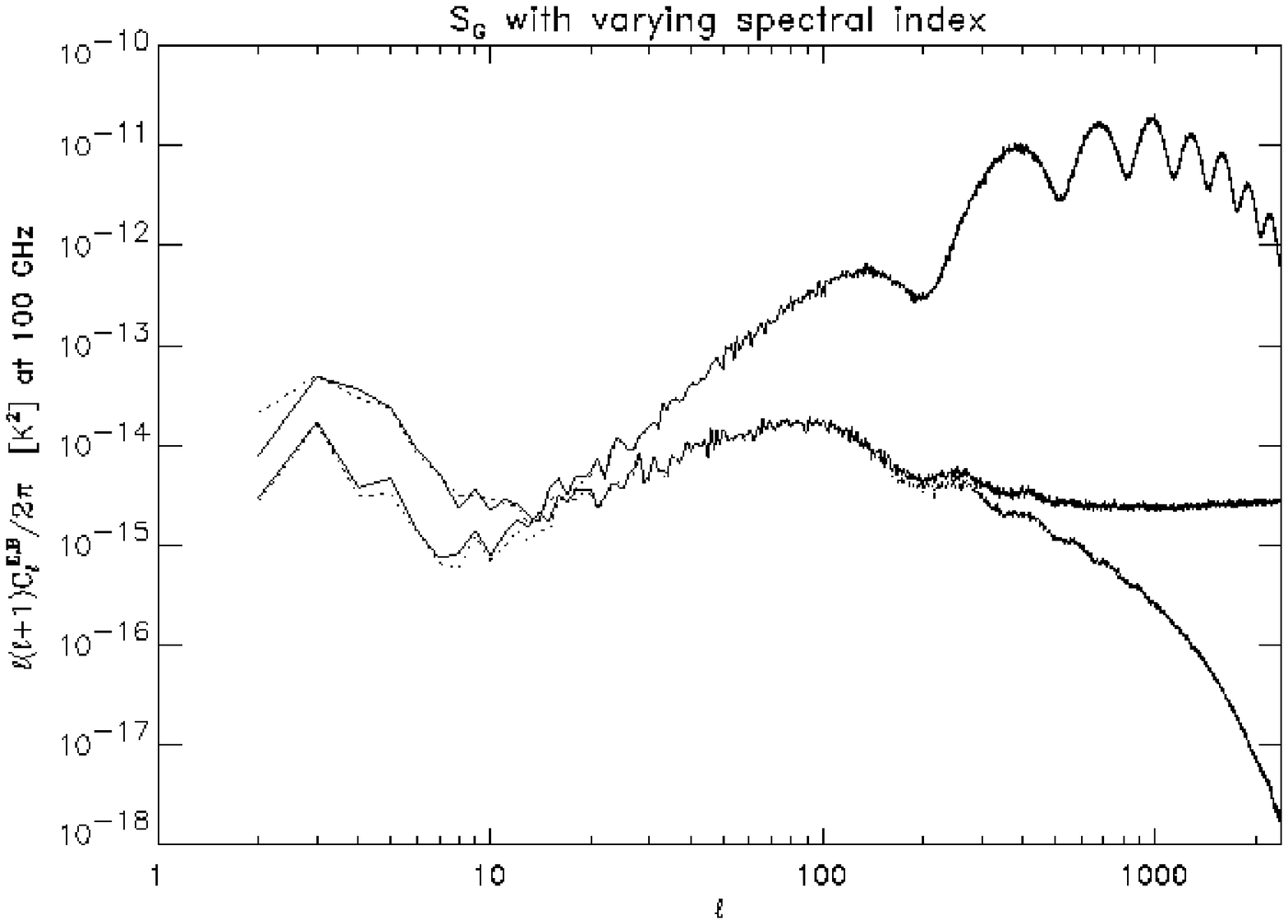,height=2.5in,width=3.in,angle=0}
\vskip .3in
\epsfig{file=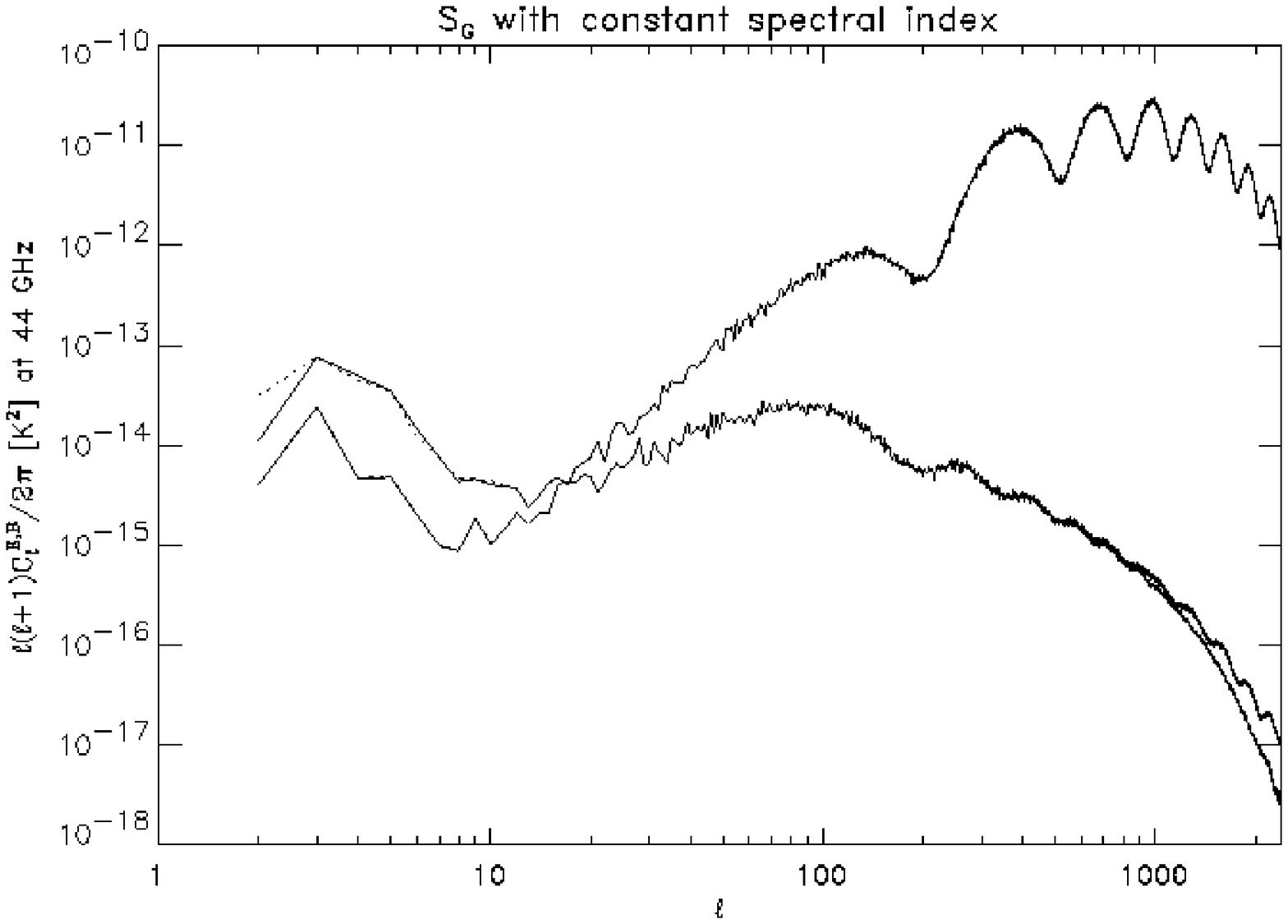,height=2.5in,width=3.in,angle=0}
\hskip .2in
\epsfig{file=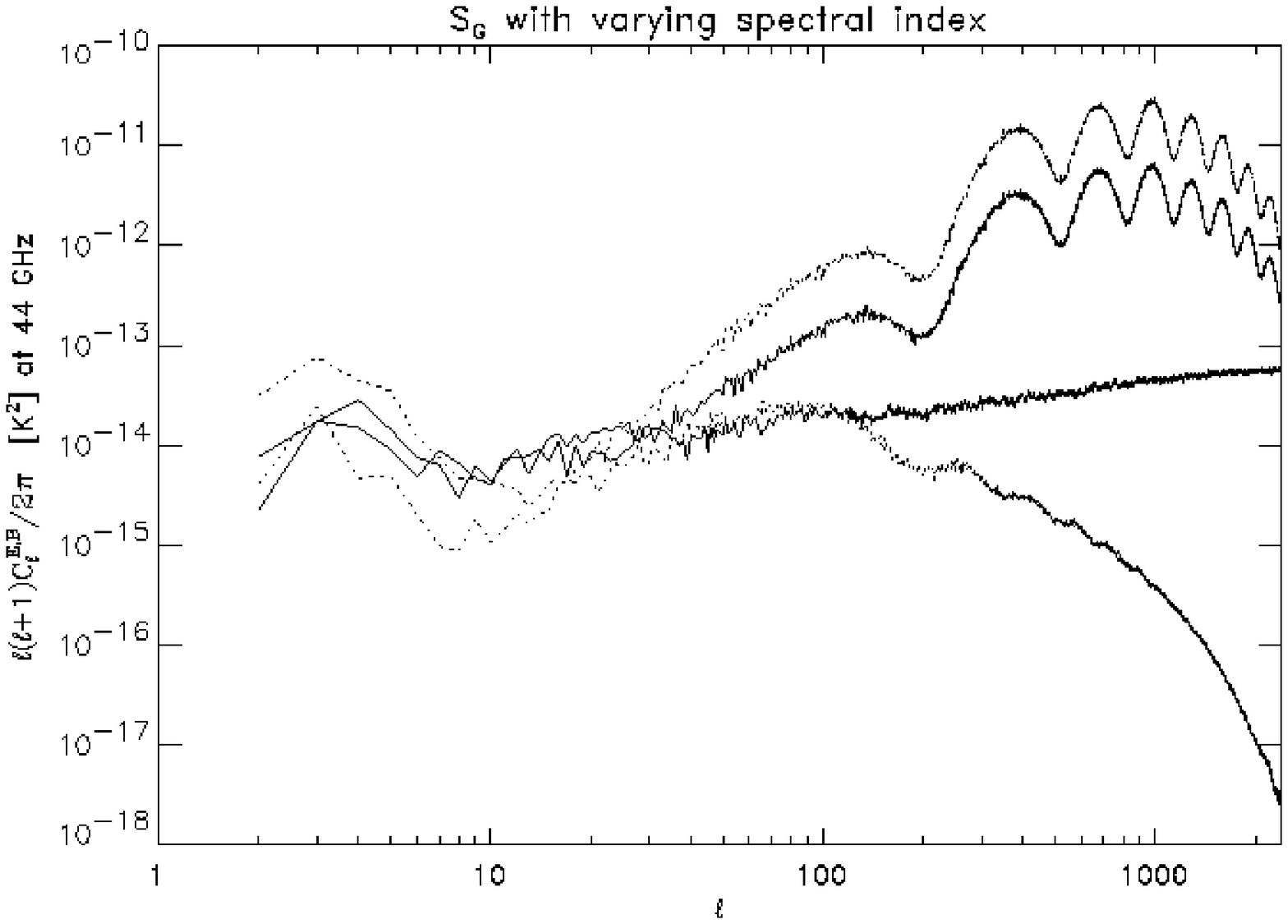,height=2.5in,width=3.in,angle=0}
\vskip .3in \caption{Original (dotted) and reconstructed (solid)
power spectra for $E$ and $B$ CMB modes in the case of $S_{G}$
model, in the absence of noise. Left panels: constant spectral
index. Right panels: space varying spectral index. Upper panels
are from inputs at 70 and 100 GHz channels,  while bottom panels
corresponds to inputs at 30 and 44 GHz. The outputs are conventionally 
plotted at the highest input frequency}
\label{nonoisemodellogiardino}
\end{center}
\end{figure*}

As we already mentioned, the {\ica} performance turns out to be stable
against relaxation of rigid frequency scalings, at least for the
spectral index variations shown in Fig.~\ref{synspecind}. In order
to illustrate quantitatively this point, we compare the quality of
the CMB reconstruction assuming either constant and varying
synchrotron spectral indices, $\beta$. In
Fig.~\ref{nonoisemodellobacci} we plot the original (dotted) and
reconstructed (solid) $C_{\ell}^{E,B}$s for CMB, in the case of
the $S_{B}$ foreground model with constant (left-hand panels) or
spatially varying $\beta$. The upper (lower) panels refer to the 
two frequency combinations 70, 100 (30, 44) GHz channels. 
Figure~\ref{nonoisemodellogiardino} shows the results of the same
analisys, but using the $S_{G}$ model.

As it can be seen, the CMB signal is well reconstructed on all
relevant scales, down to the pixel size. The same is true for the
synchrotron emission, not shown. Percent precision in frequency
scaling recovery for CMB and synchrotron is achieved (see Table
\ref{nonoisetable}). As we stressed in the previous Section, the
precision on frequency scaling recovery corresponds to the
precision on the estimation of the elements in the inverse of the
matrices ${\bf W}^{Q}$ and ${\bf W}^{U}$. Remarkably, {\ica} is
able to recover the CMB $B$ modes on all the relevant angular
scales, even if they are largely subdominant with respect to the
foreground emission, as it can be seen in Fig.~\ref{clsynsky}.
This is due to two main reasons: the difference of the underlying
statistics describing the distribution of CMB and foreground
emission, and the high angular resolution of templates ($3'.5$).
Such resolution allows the algorithm to converge close to the
right solution by exploiting the wealth of statistical information
contained in the maps. Note also that in the noiseless case with
constant synchrotron spectral index, the CMB power spectrum is
reconstructed at the same good level both for the 100, 70 GHz and
for 30, 44 GHz channel combinations, although in this frequency
range the synchrotron emission changes amplitude by a factor of
about 10. Indeed, by comparing top and bottom left panels of Figs.
\ref{nonoisemodellobacci} and \ref{nonoisemodellogiardino}, one
can note that there is only a minimum difference between the $B$
spectra at 44 GHz and at 100 GHz, arising at high $\ell$, while
the $E$ spectra exhibit no appreciable difference at all.

In the case of spatially varying spectral index (right-hand panels
of Figs.~\ref{nonoisemodellobacci} and
\ref{nonoisemodellogiardino}) a rigorous component separation is
virtually impossible, since the basic assumptions of 
rigid frequency scaling is badly violated. However {\ica} is
able to approach convergence by estimating a sort of ``mean"
foreground emission, scaling roughly with the mean value of the
spectral index distribution. Some residual synchrotron
contamination of the CMB reconstructed maps cannot be avoided,
however. This residual is proportional to the difference between
the ``true'' synchrotron emission and that corresponding to the
``mean'' spectral index and is thus less relevant at the higher
frequencies, where synchrotron emission is weaker. As shown by the
upper right-hand panels of Figs.~\ref{nonoisemodellobacci} and
\ref{nonoisemodellogiardino}, when the 70--100 GHz combination is
used, the power spectrum of the CMB $E$-mode is still well
reconstructed on all scales, and even that of the CMB $B$-mode is
recovered at least up to $\ell\simeq 100$. On smaller scales
synchrotron contamination of the $B$-mode is strong in the
$S_{G}$, but not in the $S_B$ (at least up to $\ell \simeq 1000$)
case. As expected, the separation quality degrades substantially
if the 30--44 GHz combination is used (right-hand, bottom panels
of Figs. \ref{nonoisemodellobacci} and
\ref{nonoisemodellogiardino}; see also Table \ref{nonoisetable}
where the quoted error on frequency scaling for varying spectral
index is the percentage difference between the average values
$<(\nu_{1}/\nu_{2})^{\beta}>$ computed on input and reconstructed
synchrotron maps).

\begin{table*}
\begin{center}
\caption{Percentage errors on frequency scalings reconstruction in the
noiseless case.}
\begin{tabular}{l c c c c c}
\hline
\hline
{\bf Synchrotron $S_{G}$ model} & \multicolumn{2}{c}{ \rm{\bf const.} $\beta$ }
&  \multicolumn{2}{c}{ \rm{\bf variable} $\beta$ } \\
\hline
\hline
{\rm{\bf Frequencies used (GHz)}} & 30, 44  & 70, 100   & 30, 44 & 70, 100 \\
\hline
Q synch. & 1.48 $\times 10^{-2}$ & 0.11  & 21.73  & 7.55   \\
\hline
U synch. & 5.71 $\times 10^{-3}$  & 4.56 $\times 10^{-2}$ & 21.73 & 7.55 \\
\hline
Q CMB  & 0.31 & 2.69 $\times 10^{-2}$ & 207.36  & 1.18  \\
\hline
U CMB & 0.69 & 5.83 $\times 10^{-2}$ & 222.12  & 1.21  \\
\hline
\hline
{\bf Synchrotron $S_{B}$ model} & \multicolumn{2}{c} { \rm{\bf const.}
$\beta$}
& \multicolumn{2}{c}{ \rm{\bf variable} $\beta$ } \\
\hline
\hline
{\rm{\bf Frequencies used (GHz)}} & 30, 44& 70, 100 & 30, 44 & 70, 100 \\
\hline
Q synch. & 0.18  & 1.46 & 19.19  & 1.04   \\
\hline
U synch. & 0.137  & 1.10 & 19.73 & 1.04 \\
\hline
Q CMB  & 1.01 &   8.79 $\times 10^{-2}$ & 6.15 & 9.62 $\times 10^{-2}$  \\
\hline
U CMB & 0.1448 & 1.24  $\times 10^{-2}$  & 9.028  & 0.107 \\
\hline
\hline
\end{tabular}
\label{nonoisetable}
\end{center}
\end{table*}

{To compare the {\ica} performances when $Q$ and $U$ maps are
dealt with separately or together (case $QU$), we have carried out
a Monte Carlo chain on the CMB realizations, referring to the 70
and 100 GHz channels, and computed the $rms$ error on the CMB
frequency scaling reconstruction, $\sigma_{Q}$, $\sigma_{U}$, and
$\sigma_{QU}$. The results are shown in Table
\ref{noiseless_variances}. Again the reconstruction is better when
the weaker synchrotron model $S_{B}$ is considered. The slight
difference between $\sigma_{Q}$ and $\sigma_{U}$ is probably due
to the particular realization of the synchrotron model we have
used (not changed through this Monte Carlo chain). The fact that
the difference is present for both the $S_{B}$ and $S_{G}$ models
is not surprising because the two models have different power as a
function of angular scales but have the same distribution of
polarization angles. The reason why we didn't vary the foregrounds
templates in our chain is the present poor knowledge of the
underlying signal statistics.

The separation precision when $Q$ and $U$ are considered together
is equivalent or better than if they are treated separately, as
expected since the statistical information in the maps which are
processed by {\ica} is greater. On the other hand, the fact that
$\sigma_{QU}$ is so close to $\sigma_{Q}$ and $\sigma_{U}$ clearly
indicates that for a pixel size of about $3'.5$ the statistical
information in the maps is such that the results are not greatly
improved if the pixel number is doubled. In the following we just
consider the most general case in which $Q$ and $U$ maps are
considered separately.}

\subsection{The effect of noise}
\label{the}

To study the effect of the noise on {\ica} component separation we
use a map resolution of about $7'$, corresponding to
$n_{side}=512$ in an HEALPix scheme (G\'orski et al. 1999). At
this resolution the all sky separation runs take a few seconds.
Moreover we consider only the combination of 70 and 100 GHz
channels, and a space varying synchrotron spectral index. {We
give the results for one particular noise realization and then we
show that the quoted results are representative of the typical
{\ica} performance within the present assumptions. Moreover, we
investigate how the foreground emission affects the recovered CMB
map.} The noise is assumed to be Gaussian and uniformly
distributed, with $rms$ parameterized with the signal to noise
ratio, $S/N$, where $S$ stays for CMB. As we already stressed,
noise is either subtracted during the separation process, and on
the reconstructed $C_{\ell}$, according to the estimate in
Eq,~(\ref{clnEB}). The results are affected by the residual noise
fluctuations, with power given by Eq.~(\ref{deltaclnEB}).

\begin{table*}
\begin{center}
\caption{Percentage $rms$ for CMB frequency scaling reconstruction
resulting from a Monte Carlo chain of {\ica} applied to $50$ different
CMB realizations at 70 and 100 GHz; $\sigma_{Q}$ and $\sigma_{U}$ are
obtained by treating $Q$ and $U$ separately, while $\sigma_{QU}$ is the
result when they are considered as a single array; the $1\sigma$ error on
the parameter estimation, assuming Gaussianity, is also indicated.}
\begin{tabular}{l c c c c}
\hline
\hline
\ & $\sigma_{Q}$ & $\sigma_{U}$ & $\sigma_{QU}$ \\
\hline
{\bf Synchrotron $S_{G}$ model}  & $1.23\pm 0.17$ & $1.17\pm 0.17$ & $1.21\pm 0.17$\\
\hline
{\bf Synchrotron $S_{B}$ model}  & $0.104\pm 0.015$ & $0.113\pm 0.016$ & 0$.102\pm 0.014$\\
\hline
\hline
\end{tabular}
\label{noiseless_variances}
\end{center}
\end{table*}

\begin{figure*}
\begin{center}
\epsfig{file=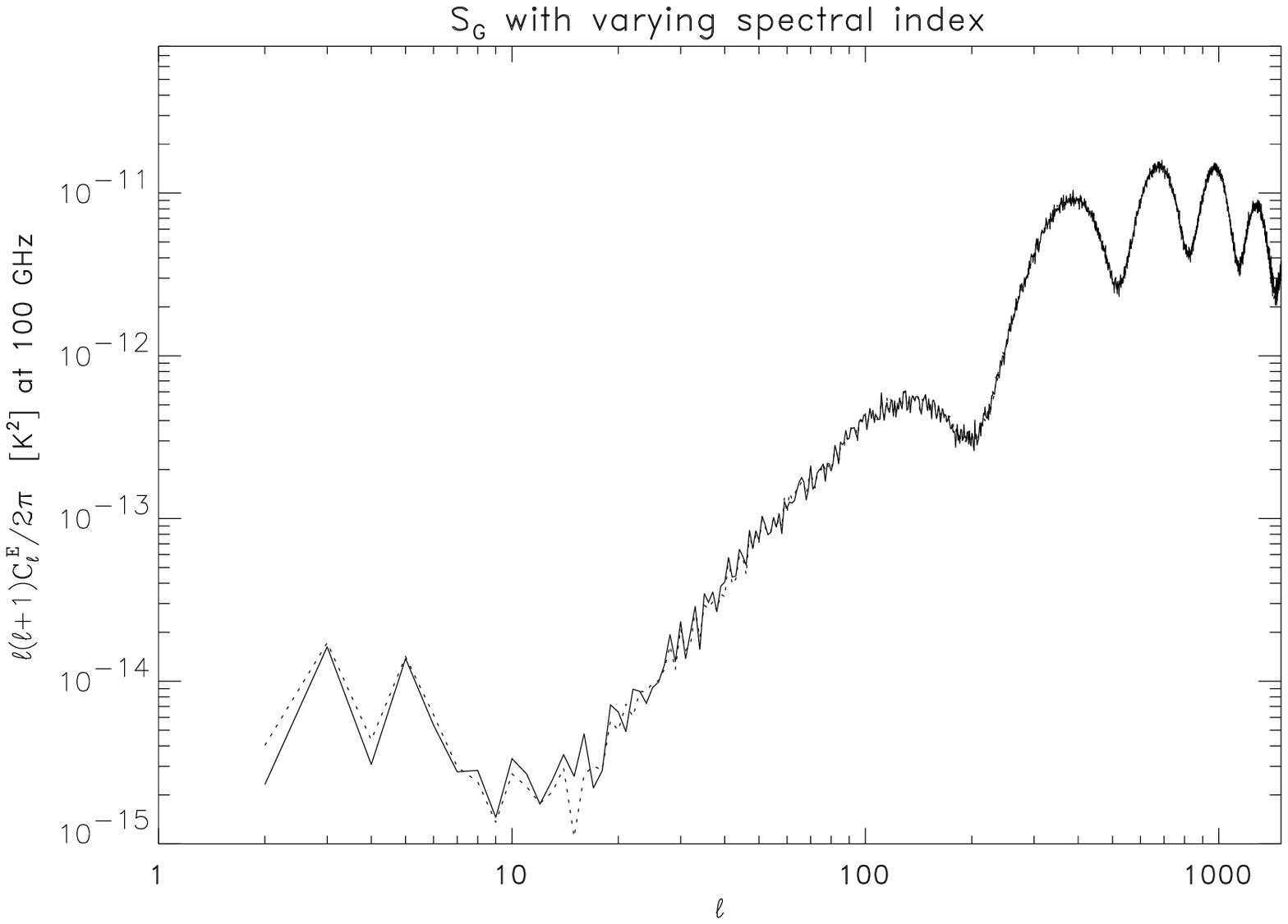,height=2.5in,width=3.in,angle=0}
\hskip .2in
\epsfig{file=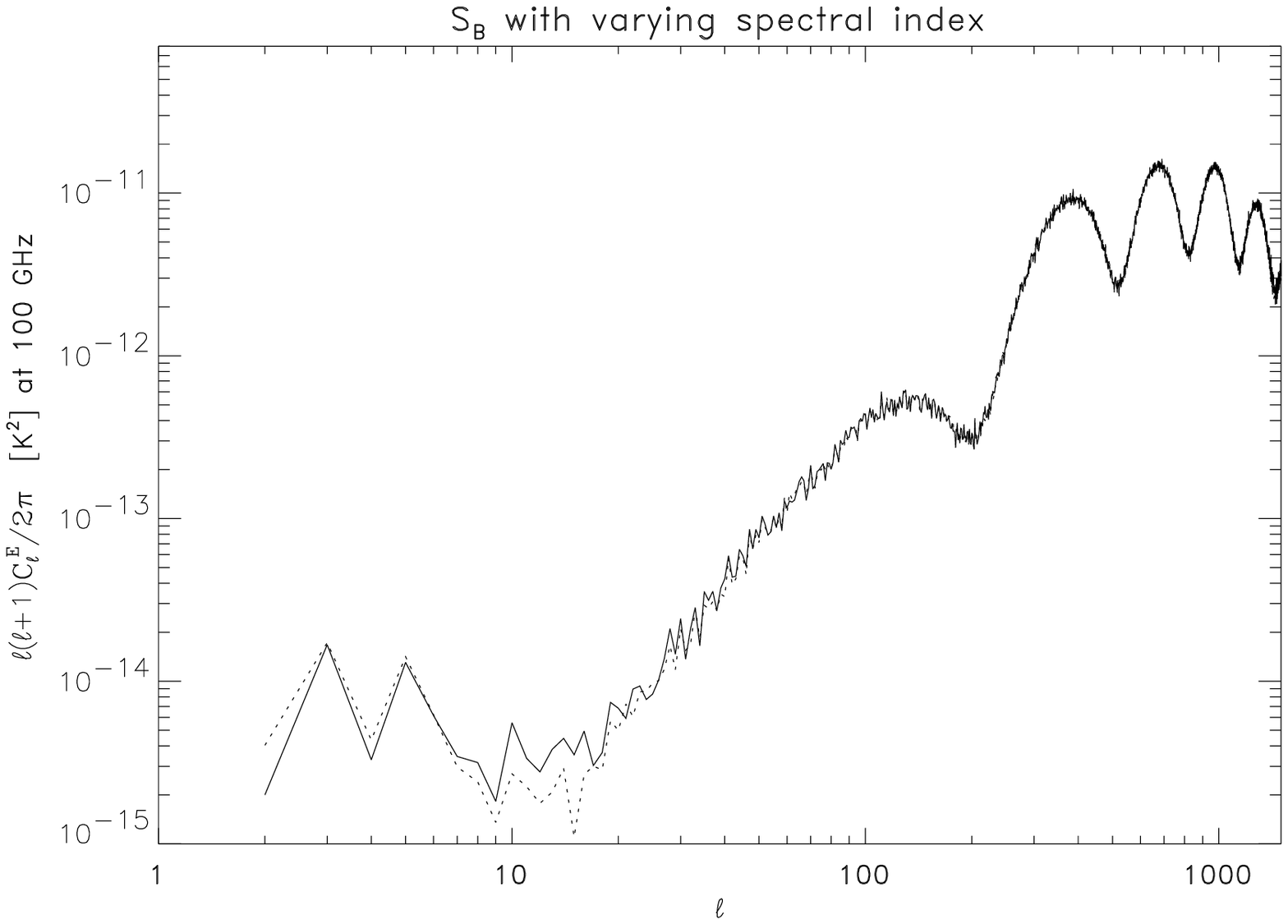,height=2.5in,width=3.in,angle=0}
\vskip .3in
\epsfig{file=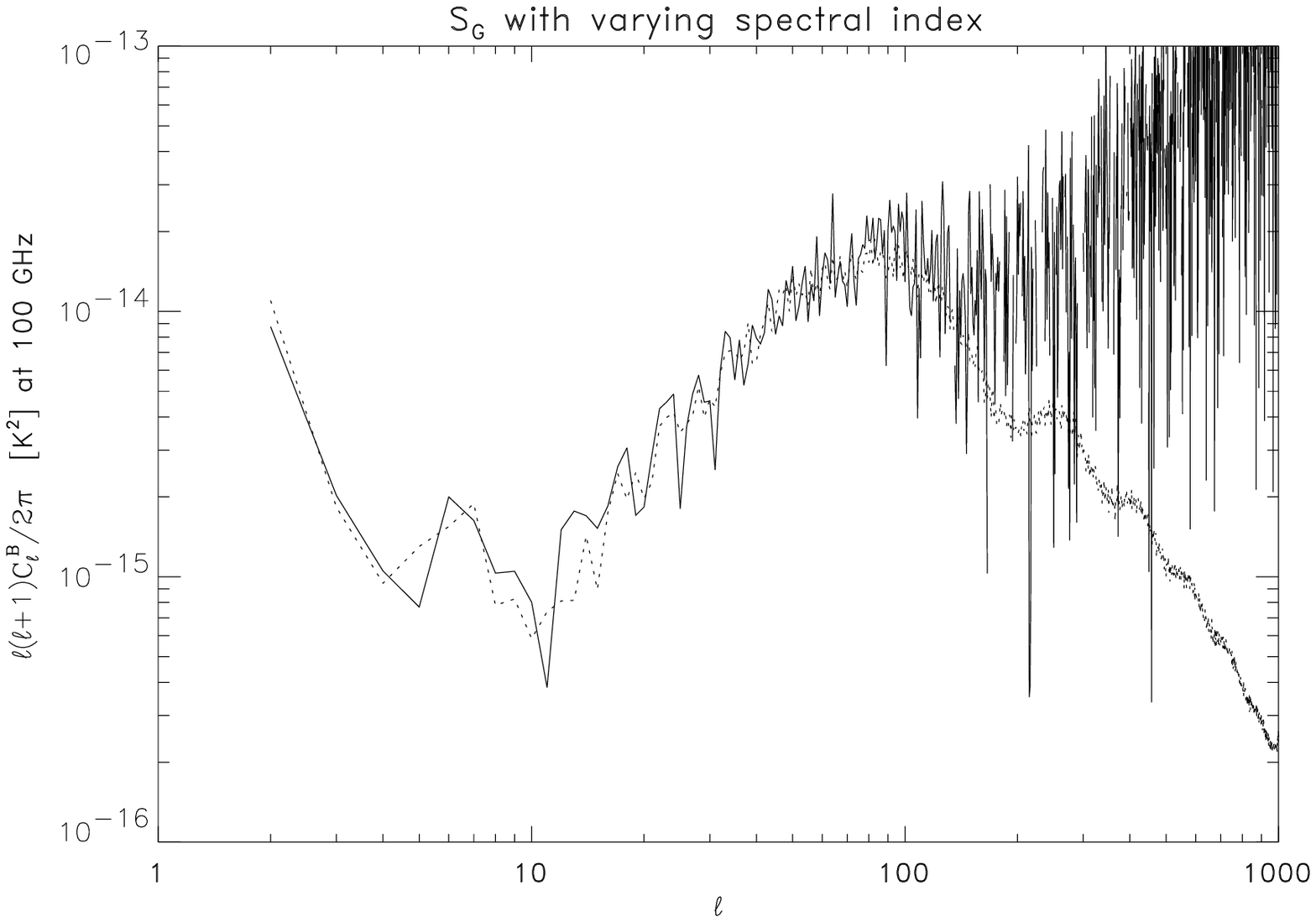,height=2.5in,width=3.in,angle=0}
\hskip .2in
\epsfig{file=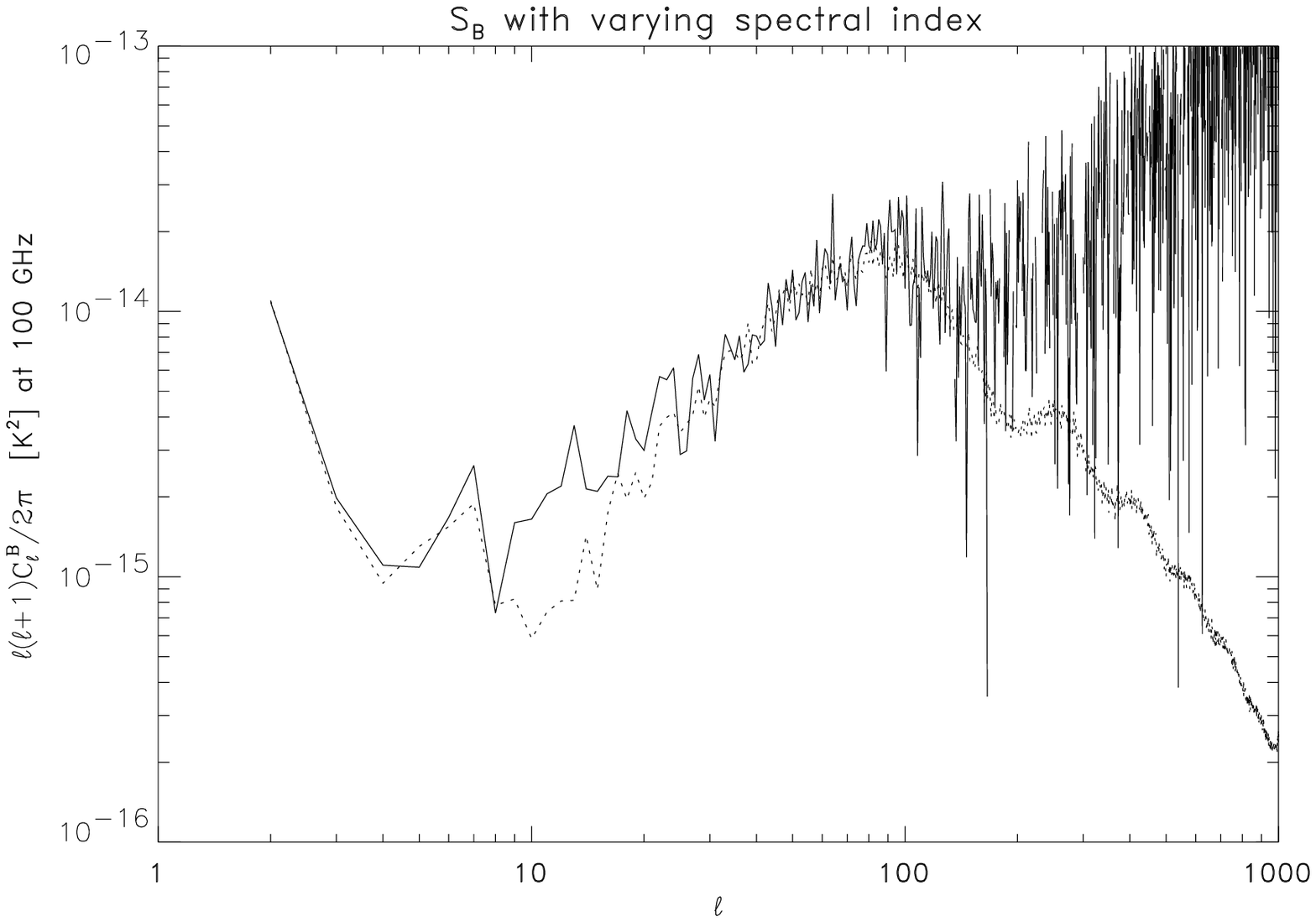,height=2.5in,width=3.in,angle=0}
\vskip .3in \caption{Original (dotted) and reconstructed (solid)
100 GHz power spectra  for $E$ (top) and $B$ (bottom) CMB modes for 
the $S_{G}$ (left) and $S_{B}$ (right) cases, assuming $S/N$=2 and
considering the 70 and 100 GHz channels; the synchrotron spectral
index is space varying.} \label{sn2}
\end{center}
\end{figure*}

As expected, the noise primarily affects the reconstruction of CMB
$B$ mode. In Fig.~\ref{sn2} we plot the reconstructed and original
CMB $E$- and $B$-mode power spectra, for $S_{G}$ (left) and
$S_{B}$ (right) foreground emission, in the case $S/N=2$. With
this level of noise, separation is still successful: the $E$-mode
power spectrum comes out very well, while that of $B$-mode is well
reconstructed up to the characteristic peak at $\ell\simeq 100$.
Table~\ref{noisetable} shows that the error on the frequency
scaling recovery, for CMB, remains, in the noisy case, at the
percent level both for $Q$ and $U$.

If the $S/N$ ratio is decreased, $B$ modes get quickly lost, while
still the algorithm is successful in recovering the $E$ power
spectrum for $S/N\gsim 0.2$ (see Fig.~\ref{sn.2.5}). The algorithm
starts failing at low multipoles, say $\ell\lsim 100$, where
synchrotron dominates over the CMB both in the $S_{G}$ and $S_{B}$
cases. The results in Fig.~\ref{sn.2.5} for the $S_{G}$ case are
averages over 8 multipoles, to avoid excessive oscillations of the
recovered spectrum. For the $S/N$ ratios in this figure, the
$B$-modes are lost on all scales. Since the $S_{G}$ model has a
higher amplitude, {\ica} is able to catch up the statistics more
efficiently than for $S_{B}$, thus being able to work with a lower
$S/N$ . Table \ref{noisetable} shows the degradation of the
separation matrix for the $S/N$ ratios of Fig.~\ref{sn.2.5},
compared to the case with $S/N=2$.

\begin{table*}
\begin{center}
\caption{Percentage errors on CMB frequency scalings
reconstruction in the noisy case by considering the 70 and 100 GHz
channels.}
\begin{tabular}{l c c c c c}
\hline
\hline
{\bf Synchrotron $S_{G}$ model} & $S/N=2$ & $S/N=0.2$ \\
\hline
Q CMB  & 1.04 & 13\\
\hline
U CMB & 0.66 & 0.89 \\
\hline
\hline
{\bf Synchrotron $S_{B}$ model} & $S/N=2$ & $S/N=0.5$ \\
\hline
Q CMB  & 0.31 & 4.26 \\
\hline
U CMB & 0.34 & 0.35 \\
\hline
\hline
\end{tabular}
\label{noisetable}
\end{center}
\end{table*}

We stress that the noise levels quoted here are not the maximum
which the algorithm can support. The quality of the separation
depends on the noise level as well as on the number of channels
considered; adding more channels, while keeping constant the
number of components to recover, generally improves the
statistical sample with which {\ica} deals and so the quality of
the reconstruction as well as the amount of noise supported. In
the next Section we show an example where a satisfactory
separation can be obtained with higher noise by considering a
combination of three frequency channels.

{We now investigate to what extent the results quoted here are
representative of the typical {\ica} performance and study how the
foreground emission biases the CMB maps recovered by {\ica}. To
this end we have performed a Monte Carlo chain of separation runs,
building for each of them a map of residuals by subtracting the
input CMB template from the recovered one and studying the
ensemble of those residual maps.

The residuals in the noiseless case are just a copy of the 
foreground emission. 
Their amplitude is greatly reduced with respect to the true 
foreground amplitude, in proportion to the accuracy of the 
recovered separation matrix. Since the latter accuracy is at the 
level of percent of better, as it can be seen in Table 
\ref{nonoisetable}, the residual foreground emission in the 
CMB recovered map is roughly the true one divided by 100. 
In terms of the angular power spectrum, see e.g. 
Fig.~\ref{clsynsky}, the residual foreground contamination 
to the CMB recovered power spectrum is roughly a factor $10^{4}$ 
less than the true one. 

In the of noisy separation a key feature is that at the present
level of architecture, the {\ica} outputs are just a linear
combination of the input channels. 
Thus, even if the separation 
goes perfect, the noise is present in the output just as the same
linear combination of the input noise templates. 
Note however that this does not mean that the noise is transmitted 
linearly to the outputs. The way the separation matrix is found depends 
non-linearly on the input data including the noise. In other words, 
the noise affects directly the estimation of the separation matrix, as 
we explained in Section \ref{instrumental}. Eq.~(\ref{sigmayQU}) 
describes only the amount of noise which affects the outputs 
after the separation matrix is found. 
As we shall see in a moment, at least in the case $S/N=2$ the main 
effect of the noise is the one given by Eq.~(\ref{sigmayQU}), dominant 
over the error induced by the noise on the separation matrix estimation. 
Moreover, the noise in the outputs reflects the input noise statistics, 
which is Gaussian and uniformly dsitributed in the sky. As we 
shall see now, this is verified if the foreground contamination is the 
stronger $S_{G}$ one. 

The results presented in Table~\ref{noise_variances} show the
ensemble average of the mean of the residuals $<\bar{r}>$ together
with its Gaussian expectation $<\bar{r}^{2}>_{G}^{1/2}$, 
and the mean $rms$ error on the CMB frequency scaling recovery, $\sigma$. 
The most important feature is that a non-zero mean value, at almost
$10\sigma$ with respect to its Gaussian expectation, is detected.
This is the only foreground contamination we find in the
residuals. 
Note that the separation matrix
precision recovery is at the percent level. That means that 
the present amount of noise does not affect significantly the accuracy 
of the separation process. Of course, if the noise is increased, the 
separation matrix estimation starts to be affected and eventually the 
foreground residual in the CMB reconstructed map will be relevant. 

We made a further check by verifying that the residuals obey a
Gaussian statistics with $rms$ given by Eq.~\ref{sigmayQU} 
on all Galactic latitudes. We constructed a map having in each 
pixel the variance built out of the 50 residual maps in our Monte Carlo 
chain. In Fig.~\ref{Gaussian_residuals} we show the $rms$ of such 
map, plus/minus the standard deviation, calculated on rings with 
constant latitude with width equal to 1 degree. 
Together with the curves built out of our Monte Carlo chain, 
we report the theoretical values according to a Gaussian 
statistics, i.e. the average given by Eq.~(\ref{sigmayQU}) equal 
to $(3.90\times 10^{-6} K)^{2}$, and the standard deviation over 
$N=50$ samples, given by 
$\sqrt{2(N-1)/N^{2}}\cdot (3.90\times 10^{-6} K)^{2}$. The agreement 
demonstrates that the Gaussian expectation is satisfied at all 
latitudes, expecially at the lowest, where the foreground contamination 
is expected to be maximum. Note that the fluctuations around the Gaussian 
theoretical levels are larger near the poles because of the enhanced sample 
variance. 

We conclude that, within the present assumptions, for a successful
separation the residual foreground contamination in the recovered
CMB map is subdominant with respect to the noise. On the other
hand, further tests are needed to check this result against a more
realistic noise model, featuring the most important systematic
effects like a non-uniform sky distribution, the presence of
non-Gaussian features etc.

\begin{figure*}
\begin{center}
\epsfig{file=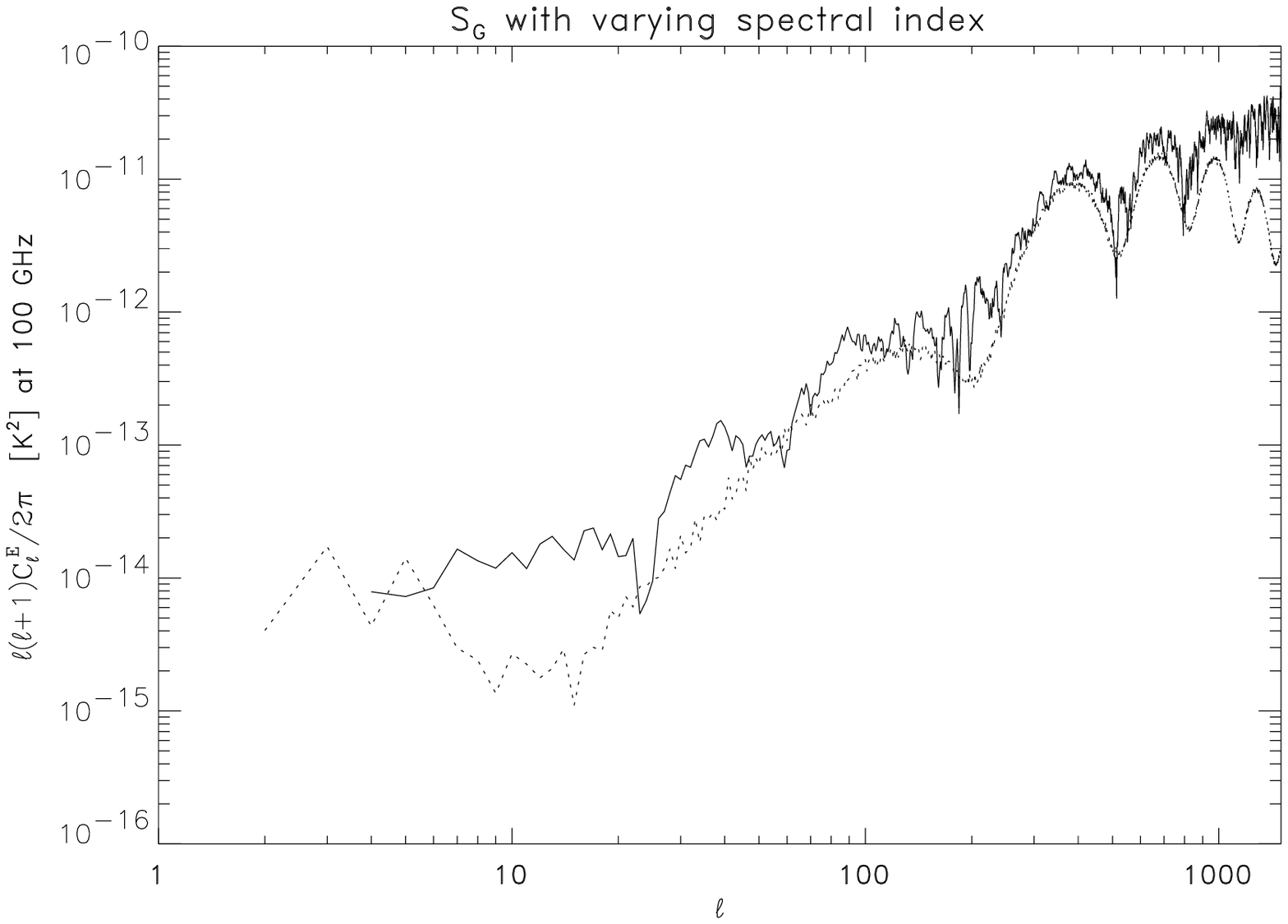,height=2.5in,width=3.in,angle=0}
\hskip .2in
\epsfig{file=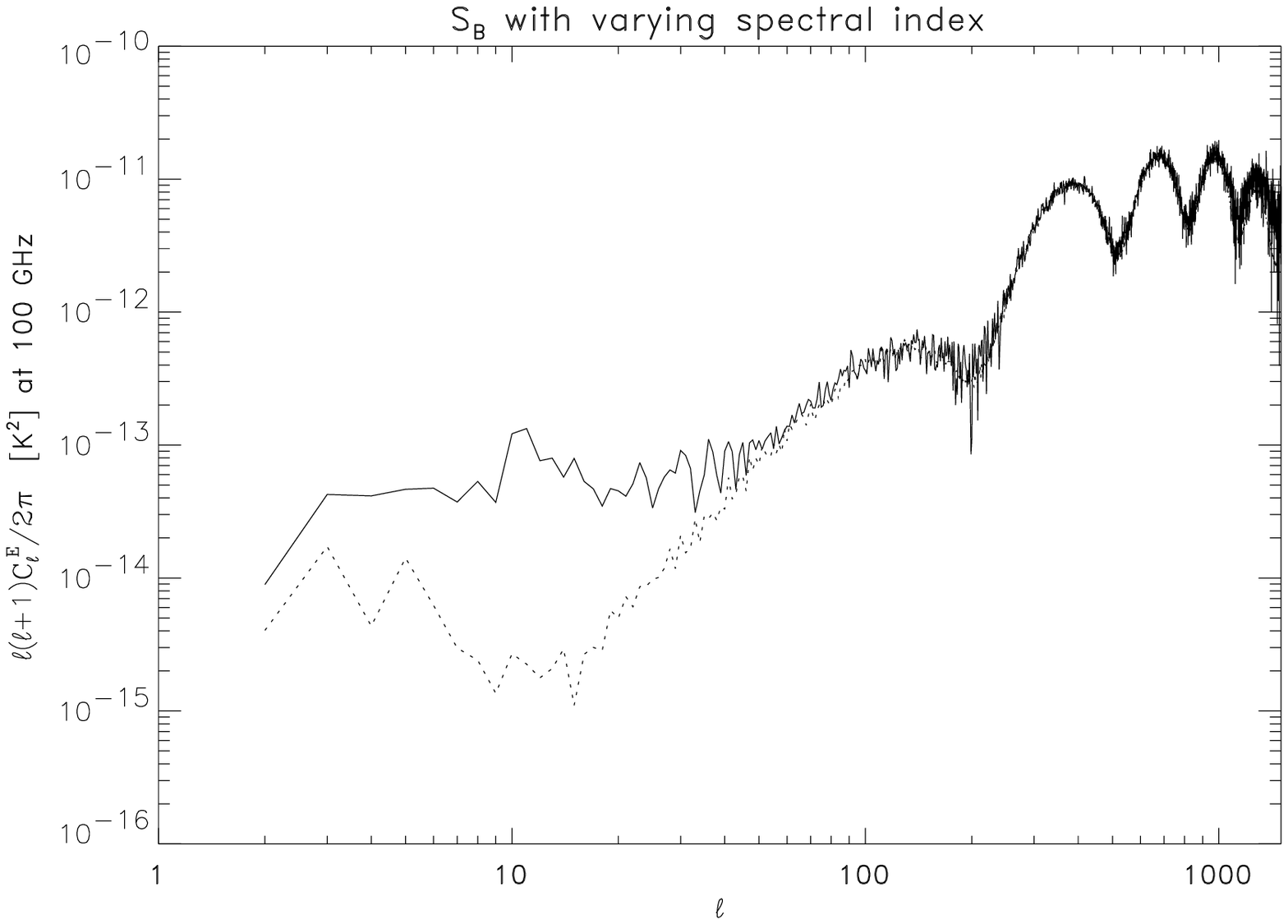,height=2.5in,width=3.in,angle=0}
\vskip .3in
\epsfig{file=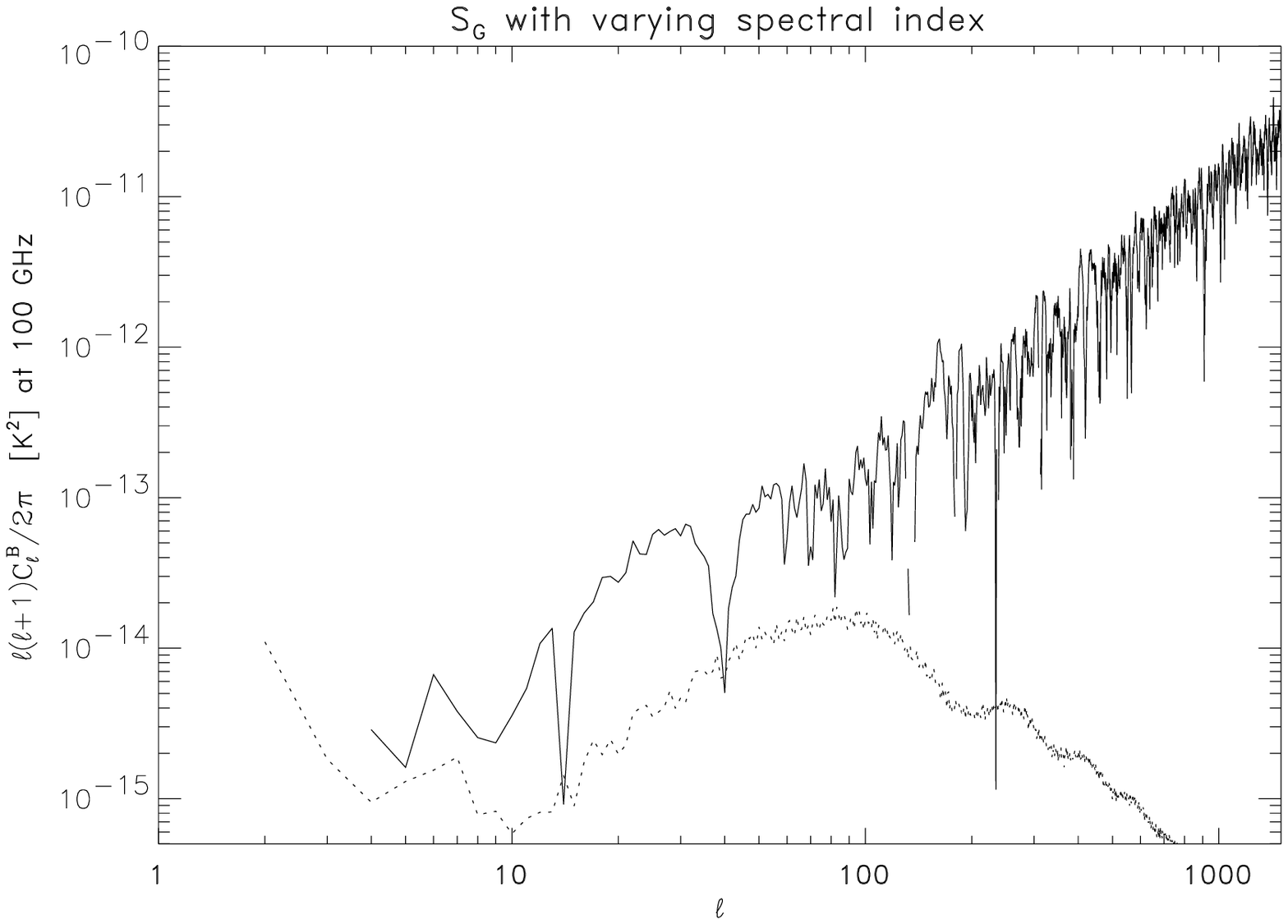,height=2.5in,width=3.in,angle=0}
\hskip .2in
\epsfig{file=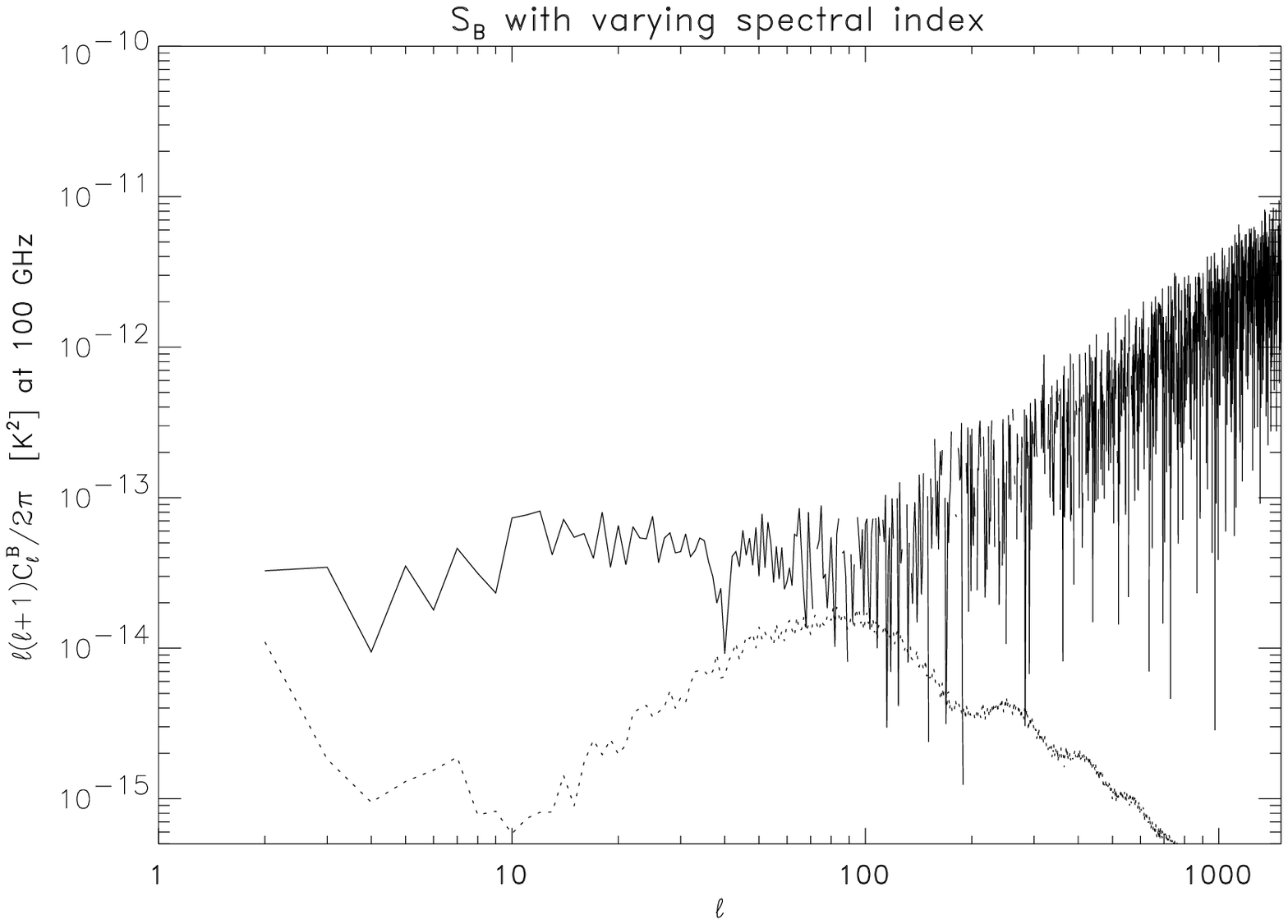,height=2.5in,width=3.in,angle=0}
\vskip .3in \caption{Input (dotted) and reconstructed (solid)
100 GHz power spectra for $E$ (top) and $B$ (bottom) CMB modes for the
$S_{G}$ (left, assuming $S/N=0.2$, averaged over 8 multipole
intervals) and $S_{B}$ (right, assuming $S/N=0.5$) cases, using
the 70 and 100 GHz channels. The synchrotron spectral index is
space varying.} \label{sn.2.5}
\end{center}
\end{figure*}

\section{An application to {\sc Planck}}
\label{application}

{In this Section, we study how {\ica} behaves in conditions
corresponding to the instrumental capabilities of {\sc Planck}.
While this work was being completed, the polarization capabilities
at 100 GHz were lost because of a funding problem of the Low
Frequency Instrument (LFI), but that capability could be restored
if the 100 GHz channel of the High Frequency Instrument (HFI) is
upgraded, as it is presently under discussion. The {\sc Planck}
polarization sensitivity in all its channels has a crucial
importance, and it is our intention to support this issue. Thus we
work assuming {\sc Planck} polarization sensitivity at 100 GHz,
highlighting the fact that of our results have been obtained under
this assumption}. At 30, 44, 70 and 100 GHz, the {\sc Planck}
beams have full width half maximum (FWHM) of 33, 23, 14 and 10
arcmin, respectively. We study the {\ica} effectiveness in
recovering $E$, $B$ and $TE$ modes, separately. We adopted, for
polarization, the nominal noise level for total intensity
measurements increased by a factor $\sqrt{2}$ (note also that due
to the 1.10 and lower HEALPix version convention to normalize 
$Q$ and $U$ following the
prescription by Kamionkowski, Kosowsky and Stebbins 1997, a
further $\sqrt{2}$ has to be taken into account when generating
$Q$ and $U$ maps out of a given power in $E$ and $B$). We neglect
all instrumental systematics in this work. The {\sc Planck}
instrumental features assumed here, with noise $rms$ in antenna
temperature calculated for a pixel size of about 3.52 arcmin
corresponding to $nside =1024$ in the HEALPix scheme, are
summarized in Table~\ref{tablePlanck}. By looking at the numbers,
it can be immediately realized that the level of noise is sensibly
higher than the one considered in the previous Section, so that
the same method would not work in this case and an improved
analysis, involving more channels as described below, is
necessary.

\begin{table*}
\begin{center}
\caption{Statistics of CMB residuals in Kelvin and percentage
errors on frequency scaling reconstruction for the case $S_{G}$,
$S/N=2$, 70 and 100 GHz channels, on 50 different noise and CMB 
realizations. The $1\sigma$ error on the parameter estimation,
assuming Gaussianity, is also indicated.}
\begin{tabular}{l c c c c c}
\hline
\hline
Stokes parameter & $<\bar{r}>$ & $<\bar{r}^{2}>_{G}^{1/2}$  & $\sigma$\\
$Q$ & $(3.94\pm 0.56)\cdot 10^{-8}$ & $(2.20\pm 0.31)\cdot 10^{-9}$ & $1.14\pm 0.16$\\
\hline
$U$ & $(4.03\pm 0.57)\cdot 10^{-8}$ & $(2.20\pm 0.31)\cdot 10^{-9}$ & $0.74\pm 0.10$\\
\hline
\hline
\end{tabular}
\label{noise_variances}
\end{center}
\end{table*}

\begin{figure*}
\begin{center}
\epsfig{file=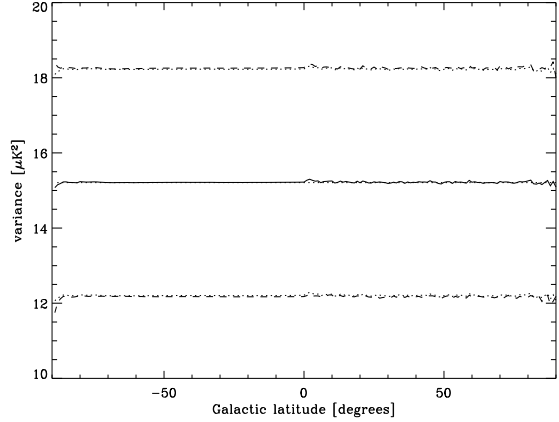,height=6.cm,width=8.cm} 
\caption{Latitude analysis of the variance map calculated out of the 
residual maps in our Monte Carlo chain, in the case $S_{G}$, $S/N=2$, 
70 and 100 GHz channels. 
The solid line is the average at the corresponding latitude, while 
the dashed curve represent the average plus/minus 
the standard deviation. The dotted lines are derived assuming Gaussian 
statistics.}
\label{Gaussian_residuals}
\end{center}
\end{figure*}

\begin{table*}
\begin{center}
\caption{{\sc Planck}-polarization performance assumed in this
work.}
\begin{tabular}{l c c c c c}
\hline
\hline
{\rm{\bf Frequency (GHz, LFI)}} & & 30 & 44 & 70 & 100 \\
\hline
FWHM (arcmin.) & & 33 & 23 & 14 & 10 \\
\hline
noise $rms$ for $3.52'$ pixels ($\mu$K) & & 80 & 81 & 69 & 52 \\
\hline
\hline
\end{tabular}
\label{tablePlanck}
\end{center}
\end{table*}

\subsection{$E$ mode}
\label{E}

Due to the high noise level, we found convenient to include in the
analysis the lower frequency channels together with those at 70
and 100 GHz. Since the {\ica} algorithm is unable to deal with
channels having different FWHM, as in Maino et al. (2002), we had
to degrade the maps, containing both signal and noise, to the
worst resolution in the channels considered. However, a
satisfactory recovery of the CMB $E$ modes, extending on all
scales up to the instrument best resolution, is still possible by
making use of the different angular scale properties of both
synchrotron and CMB. Indeed, as it can be seen in
Fig.~\ref{clsynsky}, the Galaxy is likely to be a substantial
contaminant on low multipoles, say $\ell\lsim 200$. 

For the present application, we found convenient to use a
combination of three {\sc Planck} channels, 44, 70, 100 GHz for
the $S_{G}$ and 30, 70, 100 GHz for the $S_{B}$ models,
respectively. The reason of the difference is that the $S_{B}$
contamination is weaker, and the 30 GHz channel is necessary for
{\ica} to catch synchrotron with enough accuracy. Including a
fourth channel does not imply relevant improvements. The maps,
including signals properly smoothed and noise according to the
Table~\ref{tablePlanck}, were simulated at $3.52'$ resolution,
corresponding to $n_{side}=1024$. Higher frequency maps were then
smoothed to the FWHM of the lowest frequency channel and then
re-gridded to $n_{side}=128$, corresponding to a pixel size of
about $28'$ and to a maximum multipole $\ell\simeq 400$. In all
the cases shown, the spectral index for synchrotron has been
considered variable. Fig.~\ref{clElowlGB} shows the resulting CMB
$E$ mode power spectrum after separation, for the $S_{G}$ (left)
and $S_{B}$ (right) synchrotron models. An average every 4 (left)
and 3 (right) coefficients was applied to eliminate fluctuations
going negative on the lowest signal part at $\ell\simeq 10$. The
agreement between the original spectrum and the reconstructed one
is good on all the scales probed at the present resolution, up to
$\ell\simeq 400$. The re-ionization bump is clearly visible as
well as the first polarization acoustic oscillation at $\ell\simeq
100$. Moreover, there is no evident difference in the quality of
the reconstruction between the two synchrotron models adopted.

\begin{figure*}
\begin{center}
\epsfig{file=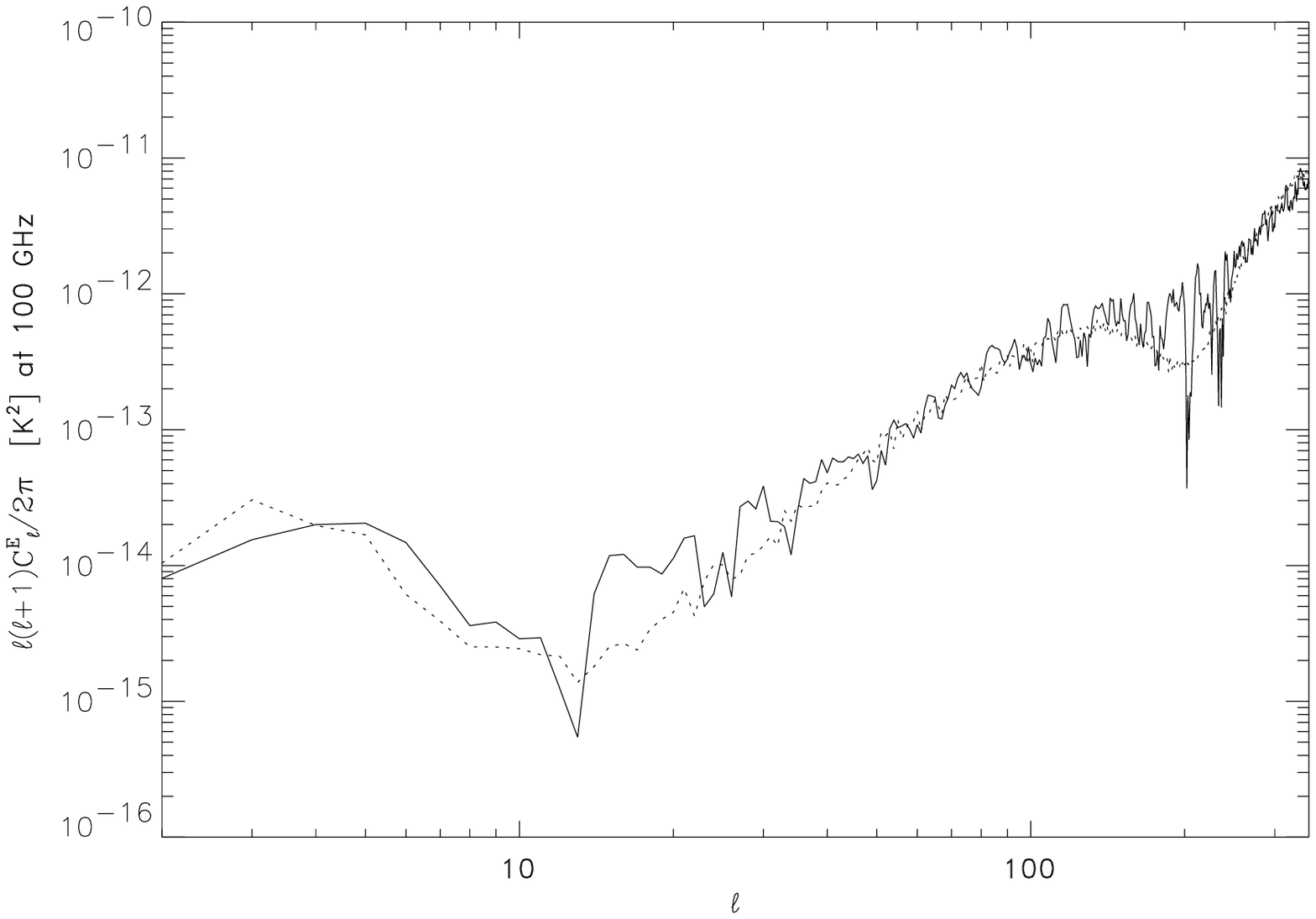,height=2.5in,width=3.in,angle=0}
\hskip .2in
\epsfig{file=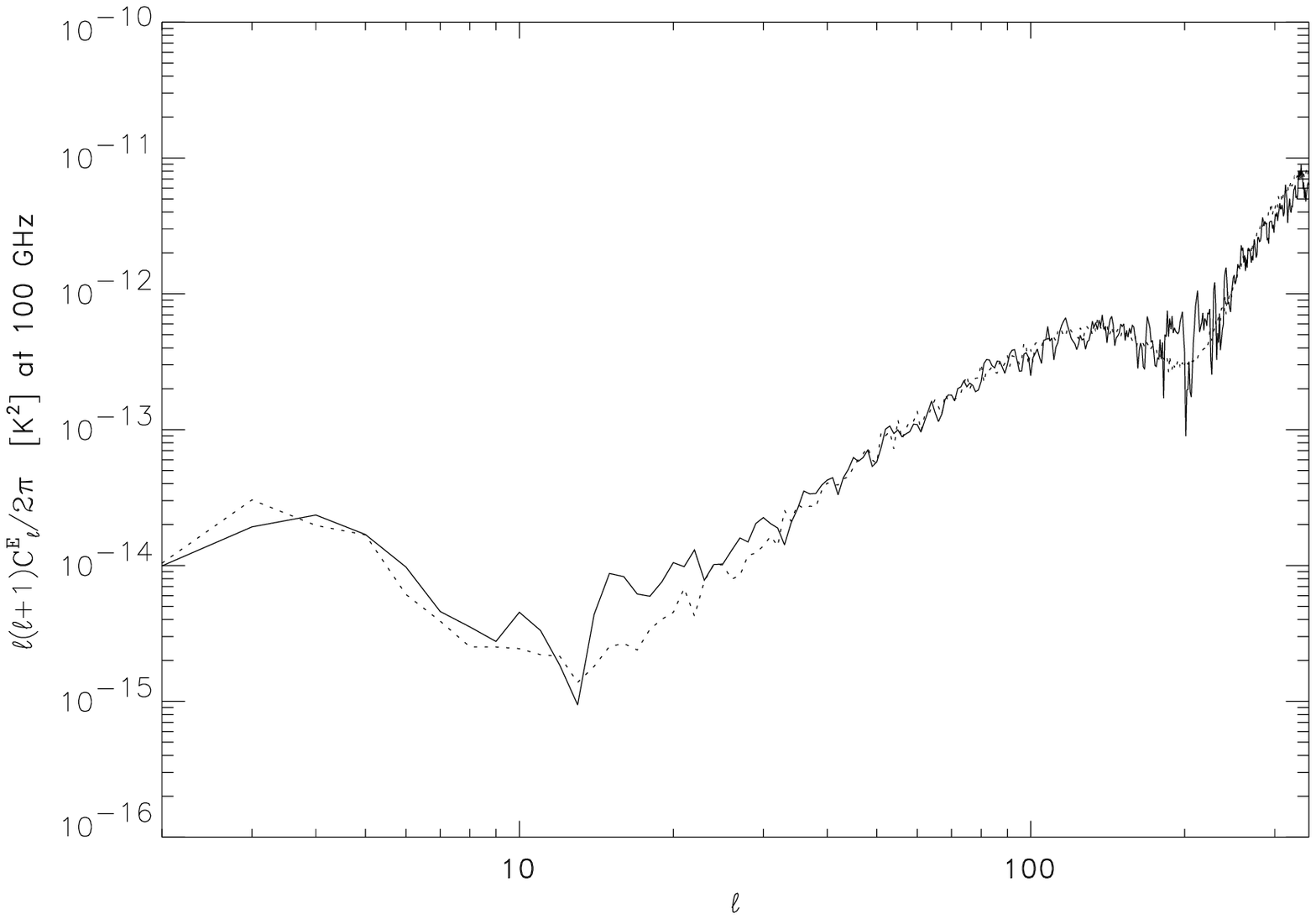,height=2.5in,width=3.in,angle=0}
\caption{Original (dotted) and reconstructed (solid) CMB
$C_{\ell}^{E}$ obtained by applying the {\ica} algorithm to the
combination of 44, 70, and 100 GHz channels for the $S_{G}$
synchrotron case (left) and of 30, 70, and 100 GHz channels for
the $S_{B}$ model (right).} \label{clElowlGB}
\end{center}
\end{figure*}

Let us turn now to the degree and sub-degree angular scales,
$\ell\gsim 200$. As we already stressed, the Galaxy is expected to
yield approximately equal power on $E$ and $B$ modes (see
Fig.~\ref{clsynsky}). On the other hand, CMB $E$ and $B$ modes are
dramatically different on sub-degree angular scales. Summarizing,
on $\ell\gsim 200$, we expect to have
\begin{equation}
C_{\ell}^{E\, Gal}\simeq C_{\ell}^{B\, Gal}
\ \ ,\ \ C_{\ell}^{E\, CMB}\gg C_{\ell}^{B\, CMB}\ .
\label{ClGalCMB}
\end{equation}
Therefore, on $\ell\gsim 200$ where the CMB contamination from
synchrotron is expected to be irrelevant, the power spectrum of
the CMB $E$ modes can be estimated by simply subtracting, together
with the noise, the $B$ power as
\begin{equation}
C_{\ell}^{E\, CMB} \simeq C_{\ell}^{E\, CMB+Gal}-C_{\ell}^{B\, CMB+Gal}-
C_{\ell}^{E\, noise}\ ,
\label{smart}
\end{equation}
where the total map CMB$+$Gal is used without any separation
procedure. In other words, there is no need to perform separation
to get the CMB $E$ modes at high multipoles, because they are
simply obtained by subtracting the $B$ modes of the sky maps,
since the latter are dominated by synchrotron which has almost
equal power on $E$ and $B$ modes. Fig.~\ref{clEhighlGB} shows the
results of this technique applied to the {\sc Planck}-HFI 100 GHz
channel, assumed to have polarization capabilities, for both the
$S_{G}$ and $S_{B}$ synchrotron models. Residual fluctuations are
higher in the $S_{G}$ case since the synchrotron contamination is
stronger. In both cases, CMB $E$ modes are successfully recovered
in the whole interval $100\lsim\ell\lsim 1000$. It has also to be
noted that the same subtraction technique would not help on the
lower multipoles considered before, since in that case the
foreground contamination is so strong that the tiny fluctuations
making Galactic $E$ and $B$ modes different are likely to hide the
CMB signal anyway.
\begin{figure*}
\begin{center}
\epsfig{file=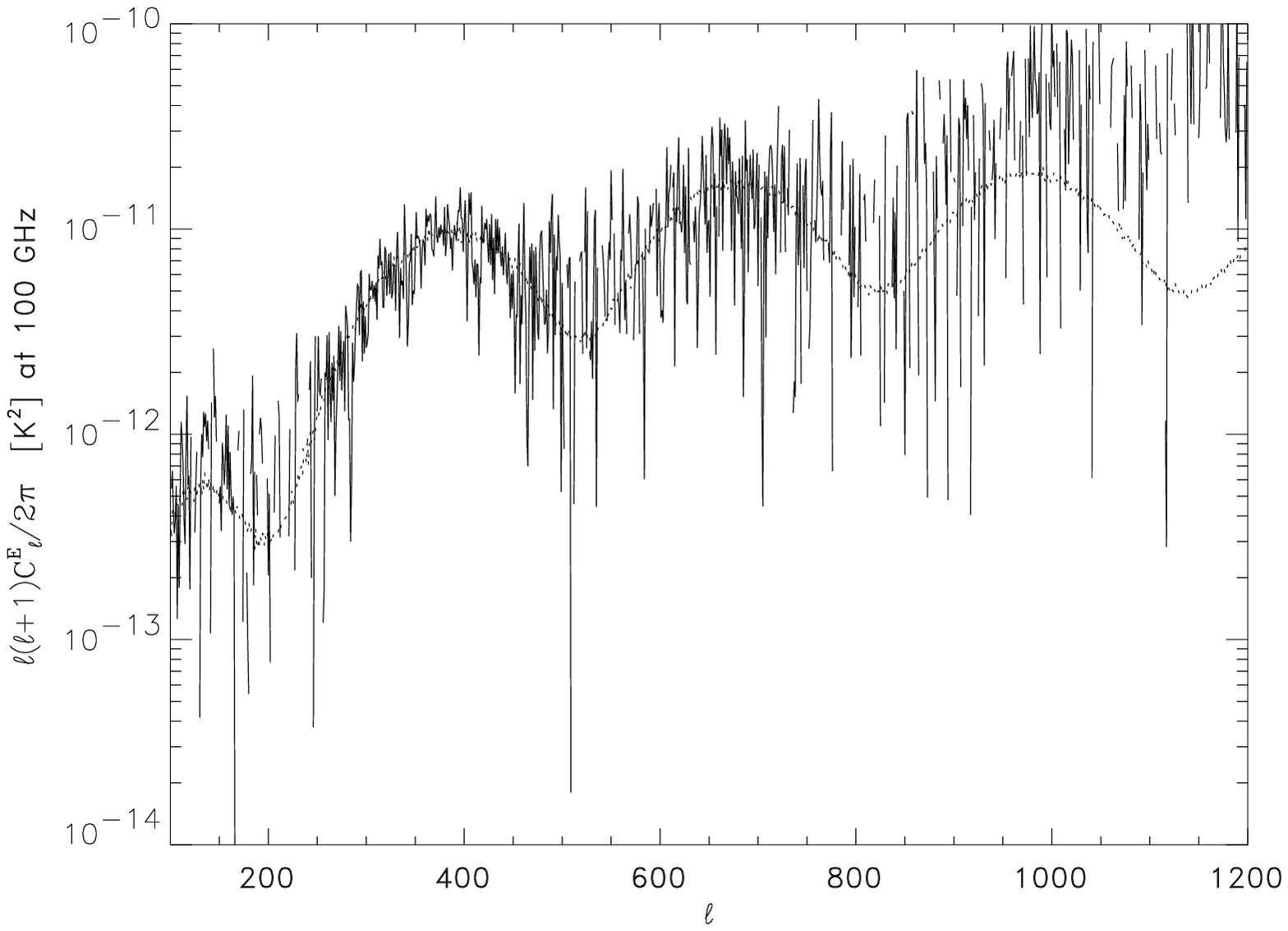,height=2.5in,width=3.in,angle=0}
\hskip .2in
\epsfig{file=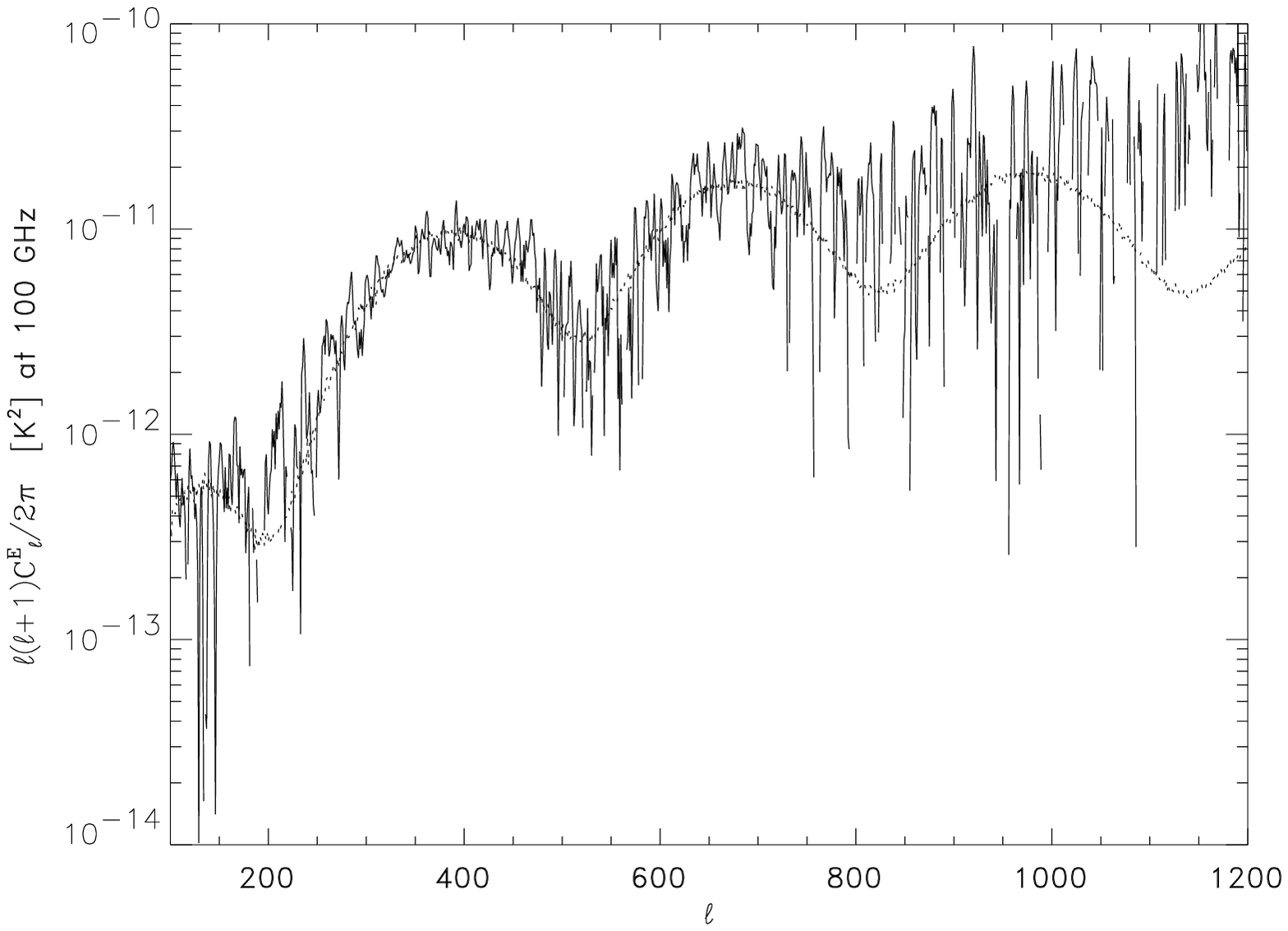,height=2.5in,width=3.in,angle=0}
\caption{Input (dotted) and reconstructed (solid) CMB
$C_{\ell}^{E}$ obtained by subtracting the expected level of noise
as well as the synchrotron contaminations $S_{G}$ (left) and
$S_{B}$ (right) assumed to be matched by the $B$-mode map. The
adopted instrumental capabilities are those of the {\sc
Planck}-HFI channel at 100 GHz, assumed to be
polarization-sensitive.} \label{clEhighlGB}
\end{center}
\end{figure*}

Our results here can be summarized as follows: in the case of {\sc
Planck} capabilities, the {\ica} technique makes it possible to
remove substantially the foreground contamination in the regions
in which that is expected to be relevant. {\sc Planck} is likely
to measure the CMB $E$ modes over all multipoles up to $\ell\simeq
1000$.

\subsection{$B$ mode}
\label{B}


In Fig.~\ref{clBlowlGB} the $B$-mode power spectrum after {\ica}
separation described in the previous section is shown for the
$S_{G}$ (left) and $S_{B}$ (right) synchrotron models. An average
over 13 multipoles has been applied to both cases in order to
avoid fluctuations going negative. The reconstructed signal
approaches the original one at very low multipoles, say $\ell\lsim
5$. At higher multipoles, where the $B$ signal is generated by
gravitational waves, the overall amplitude appears to be
recovered, even if with major contaminations especially in the
region where the signal is low, i.e. right between the
re-ionization bump and the raise toward the peak at $\ell\simeq
100$. Needless to say, such contaminations are due to a residual
foreground emission.

In the insets of Fig.~\ref{clBlowlGB} we show (data points) the
recovered B-mode power spectrum in the range between $30\le\ell\le
120$, averaged over 20 multipoles, with error bars given by
Eq.~(\ref{deltaclnEB}). Even if the contamination is substantial,
especially for $\ell\ge 100$ and for the $S_{G}$ case, the results
show a sign of the characteristic rise of the spectrum due to
cosmological gravitational waves.

Concluding, our results indicate that the {\ica} technique is able
to remove substantially the foreground contamination of the $B$
mode, up to the peak at $\ell\lsim 100$ if the tensor to scalar
perturbation ratio is at least $30\%$.
\begin{figure*}
\begin{center}
\epsfig{file=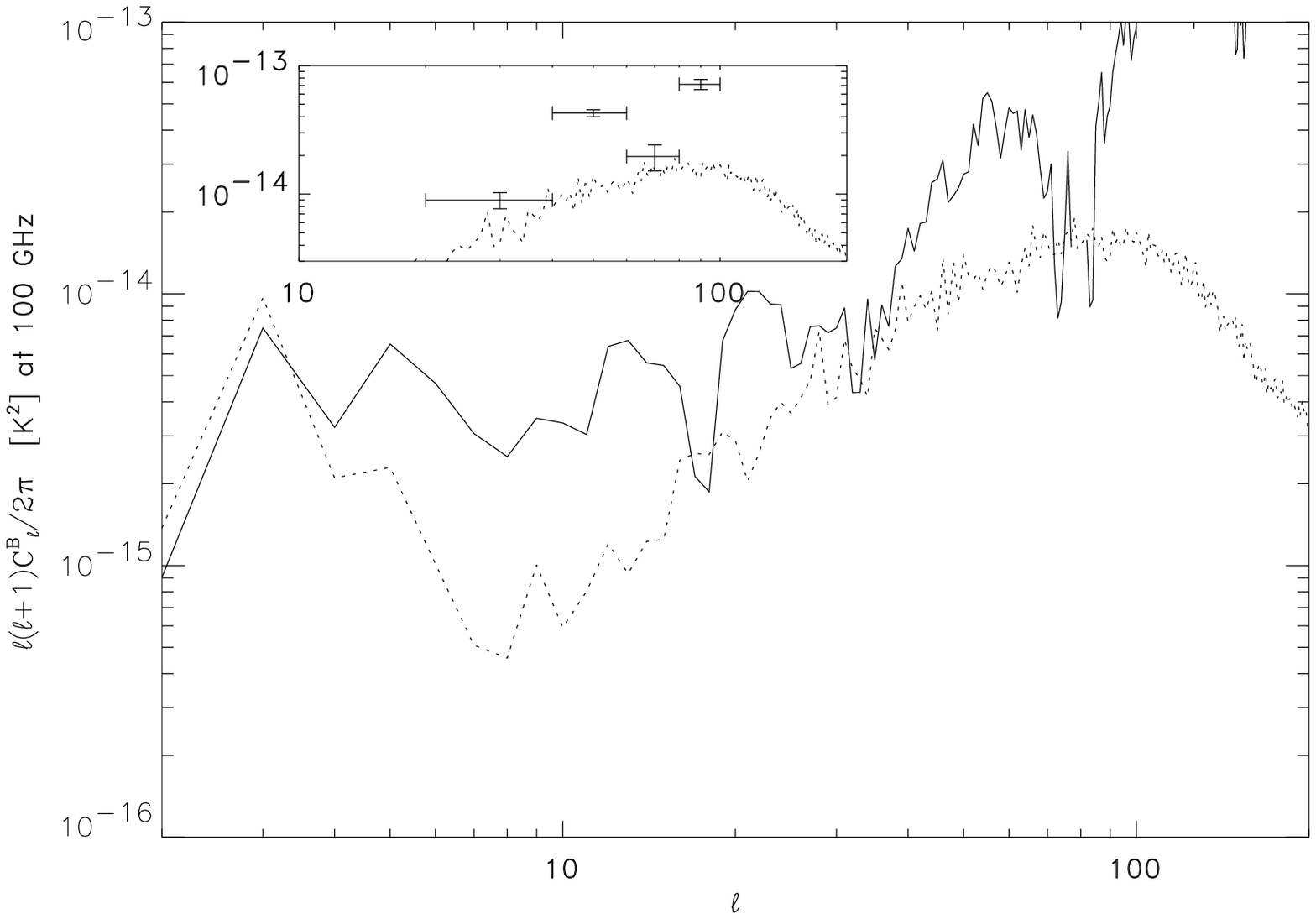,height=2.5in,width=3.in,angle=0}
\hskip .2in
\epsfig{file=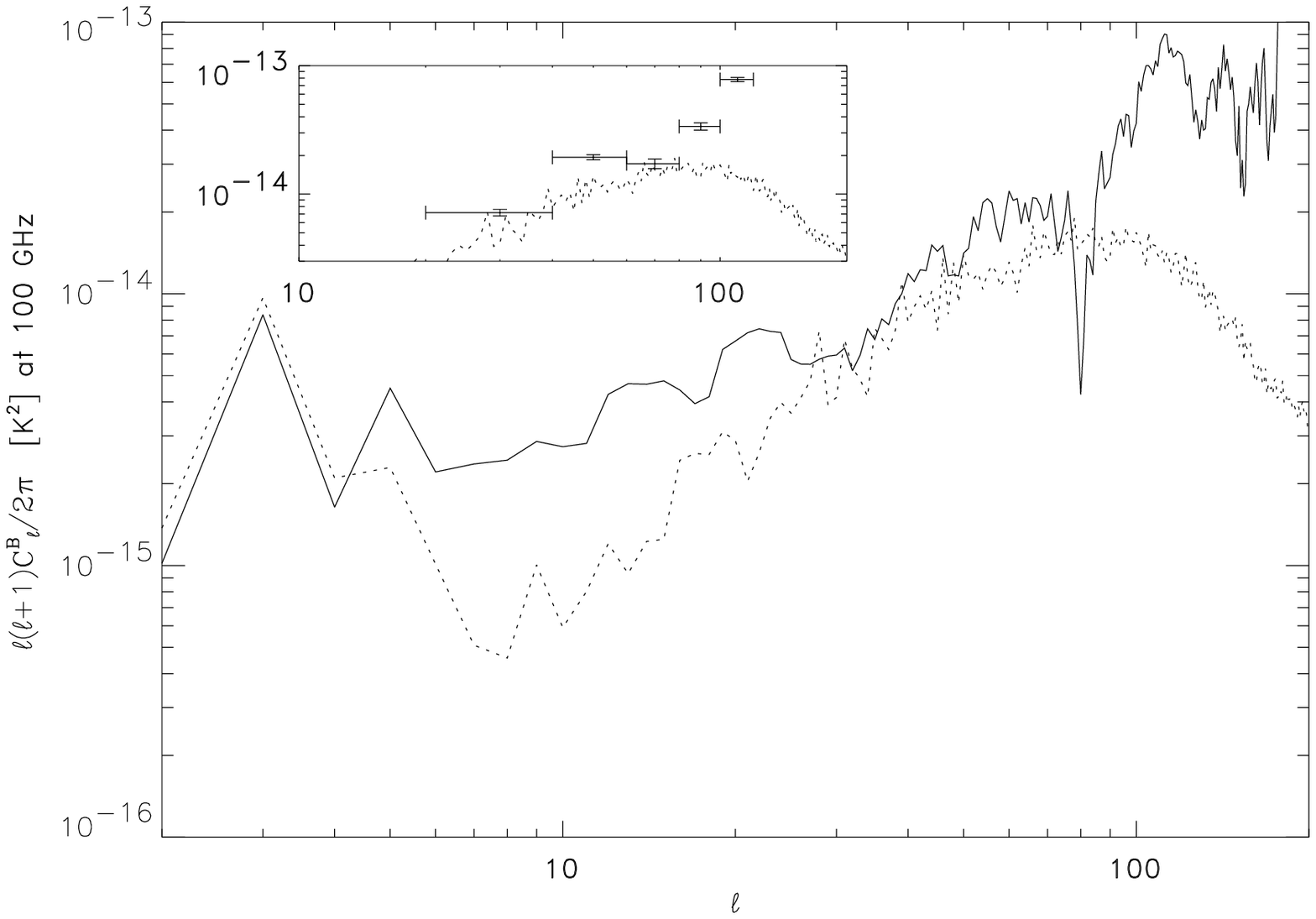,height=2.5in,width=3.in,angle=0}
\caption{Original (dotted) and reconstructed (solid) CMB
$C_{\ell}^{B}$ obtained by applying the {\ica} algorithm to the
combination of {\sc Planck} 44, 70, and 100 GHz channels for the
$S_{G}$ synchrotron case (left) and to the combination of 30, 70,
and 100 GHz channels for the $S_{B}$ model (right). The data
points in the insets show the recovered $B$-mode power spectrum in
the range $30\le\ell\le 120$ averaged over 20 multipoles; the
error bars are given by Eq.~(\ref{deltaclnEB}).} \label{clBlowlGB}
\end{center}
\end{figure*}

\subsection{$TE$ mode}
\label{TE}

\begin{figure*}
\begin{center}
\epsfig{file=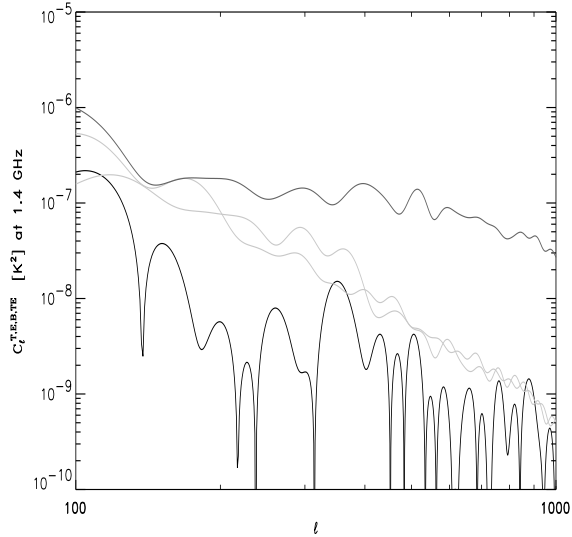,height=3.in,width=3.in,angle=0}
\caption{Power spectra of synchrotron $T$ (solid line),$E$ and $B$
(light lines), and $TE$ (heavy solid line) modes in the fan region
at medium Galactic latitudes, at 1.4 GHz (Uyaniker et al. 1999).}
\label{fuya}
\end{center}
\end{figure*}

While the cosmological $TE$ power spectrum is substantially
stronger than that of any other polarized CMB mode, the opposite
should happen in the case of foregrounds. On degree and sub-degree
angular scales, a measure of the synchrotron $TE$ power spectrum
can be achieved in the radio band. In the Parkes data at 1.4 GHz,
Uyaniker et al. (1999) were able to isolate a region exhibiting
pretty low rotation measures, named ``fan region", which is
therefore expected to be only weakly affected by Faraday
depolarization. This and other regions from the existing surveys
in the radio band were used to predict the synchrotron power for
the $S_{B}$ scenario (Baccigalupi et al. 2001).

In Fig. \ref{fuya} we show the $T$, $E$, $B$, and $TE$ power
spectra for the fan region. Total intensity anisotropies are
represented by the upper curve (solid). $E$ and $B$ modes (light
lines) have very similar behavior. The $TE$ mode (heavy solid
line) is the weakest and, as it can be easily seen by scaling the
$TE$ amplitude in Fig. \ref{fuya} with the typical spectra index
for synchrotron, it is markedly below the expected cosmological
$TE$ signal at CMB frequencies. Both synchrotron models $S_{G}$
and $S_{B}$, are consistent with this result as illustrated by in
Fig. \ref{TESYNCMB}. It is straightforward to check that our
models for the synchrotron emission have a $TE$ power spectrum not
far from the one in Fig. \ref{fuya}, when scaled to the
appropriate frequency.
\begin{figure*}
\begin{center}
\epsfig{file=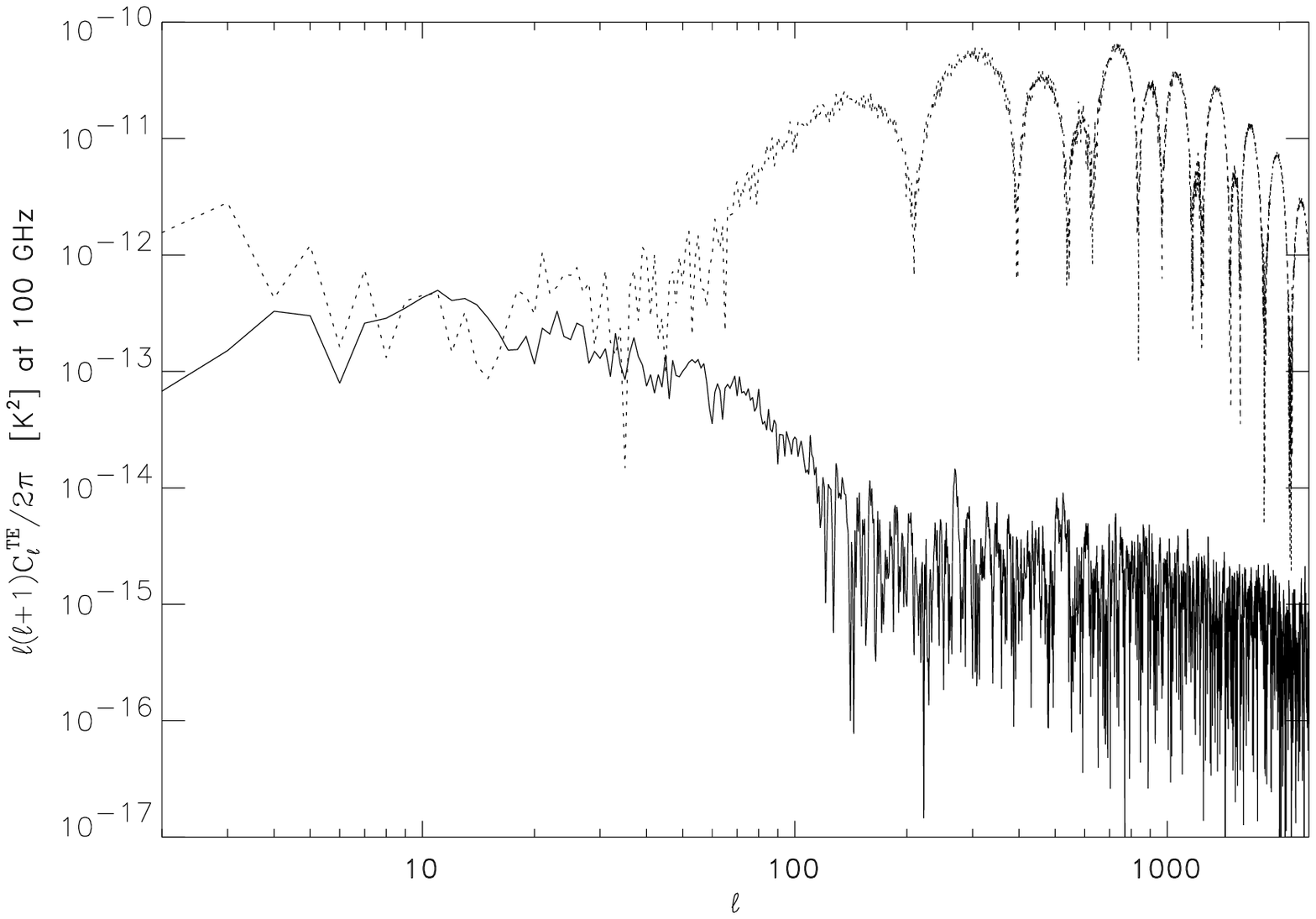,height=2.5in,width=3.in,angle=0}
\hskip .2in
\epsfig{file=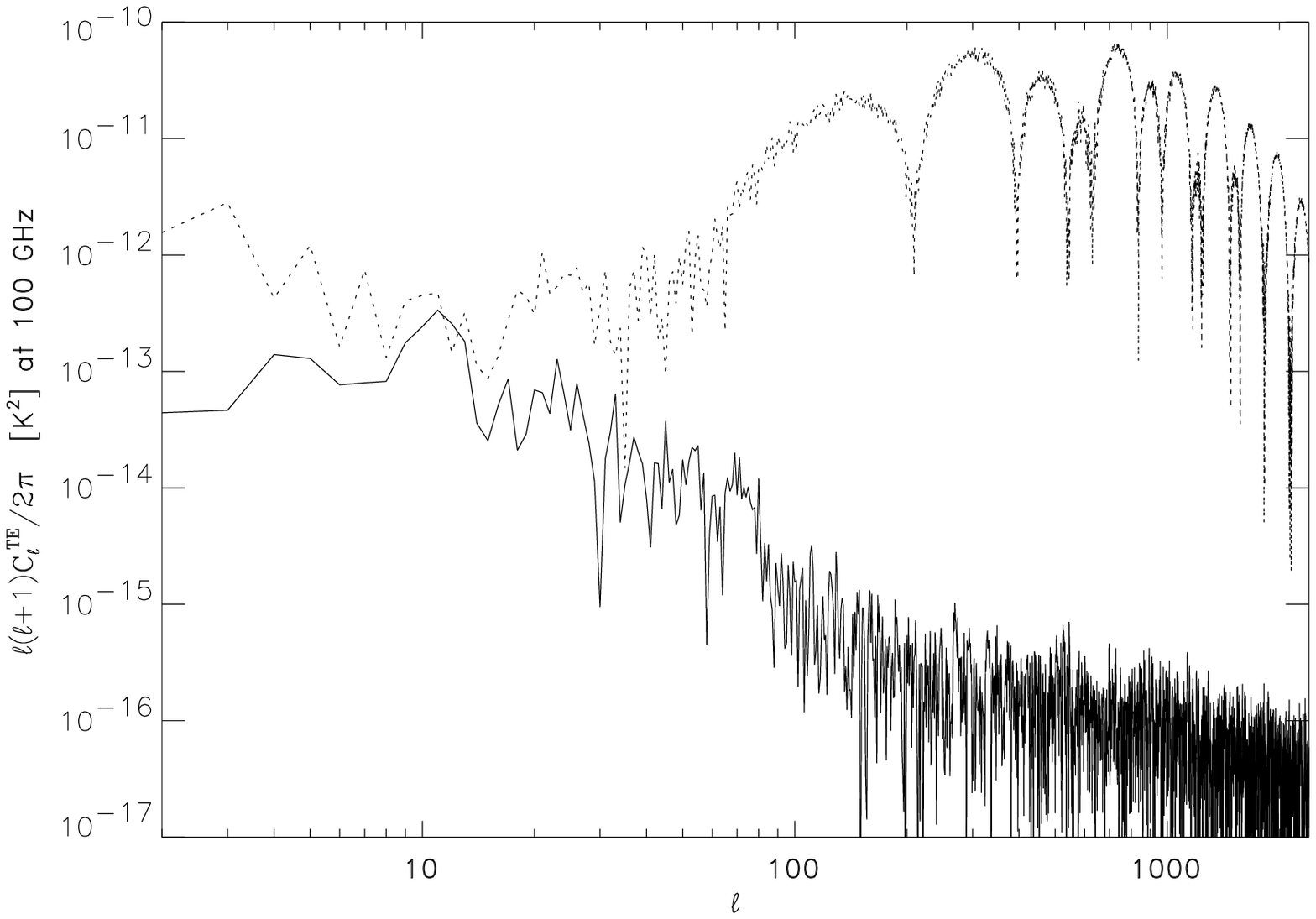,height=2.5in,width=3.in,angle=0}
\caption{$C_{\ell}^{TE}$ of CMB (dotted) compared with that of
synchrotron according to the $S_{G}$ (solid, left) and $S_{B}$
(solid, right) models, at 100 GHz.} \label{TESYNCMB}
\end{center}
\end{figure*}

From the point of view of CMB observations, this means that, if
the synchrotron contamination at microwave frequencies is well
represented by its signal in the radio band, at least on degree
and sub-degree angular scales the contamination from synchrotron
is almost absent due to the change in the magnetic field
orientation along the line of sight. On the other hand, on larger
scales, as it can be seen in Fig. \ref{TESYNCMB}, the
contamination could be relevant both in the $S_{G}$ and $S_{B}$
cases and we perform component separation as described in Section
\ref{E} for the {\sc Planck} case. In Fig. \ref{TElowl} we show
the recovery of the CMB $TE$ mode, obtained by combining the
templates of $Q$ and $U$ maps obtained after {\ica} application as
in Section \ref{E}, with the CMB $T$ template obtained, still with
{\ica} based component separation strategy, in Maino et al.
(2002). Oscillations due to residual noise are visible in the
recovered $C_{\ell}^{TE}$. However, as in the case of the
$E$-mode, the procedure was successful in substantially removing
the contamination.
\begin{figure*}
\begin{center}
\epsfig{file=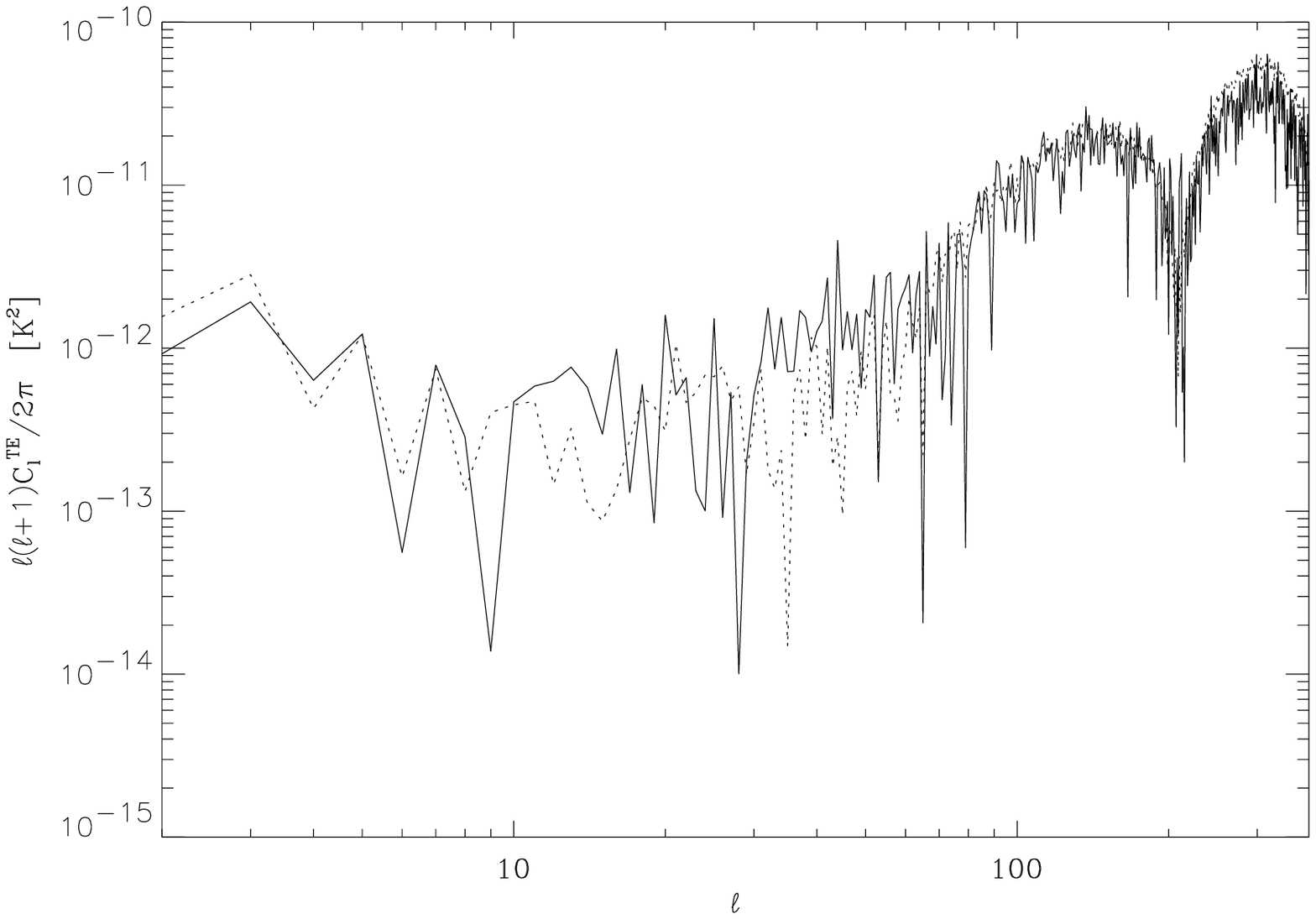,height=2.5in,width=3.in,angle=0}
\hskip .2in
\epsfig{file=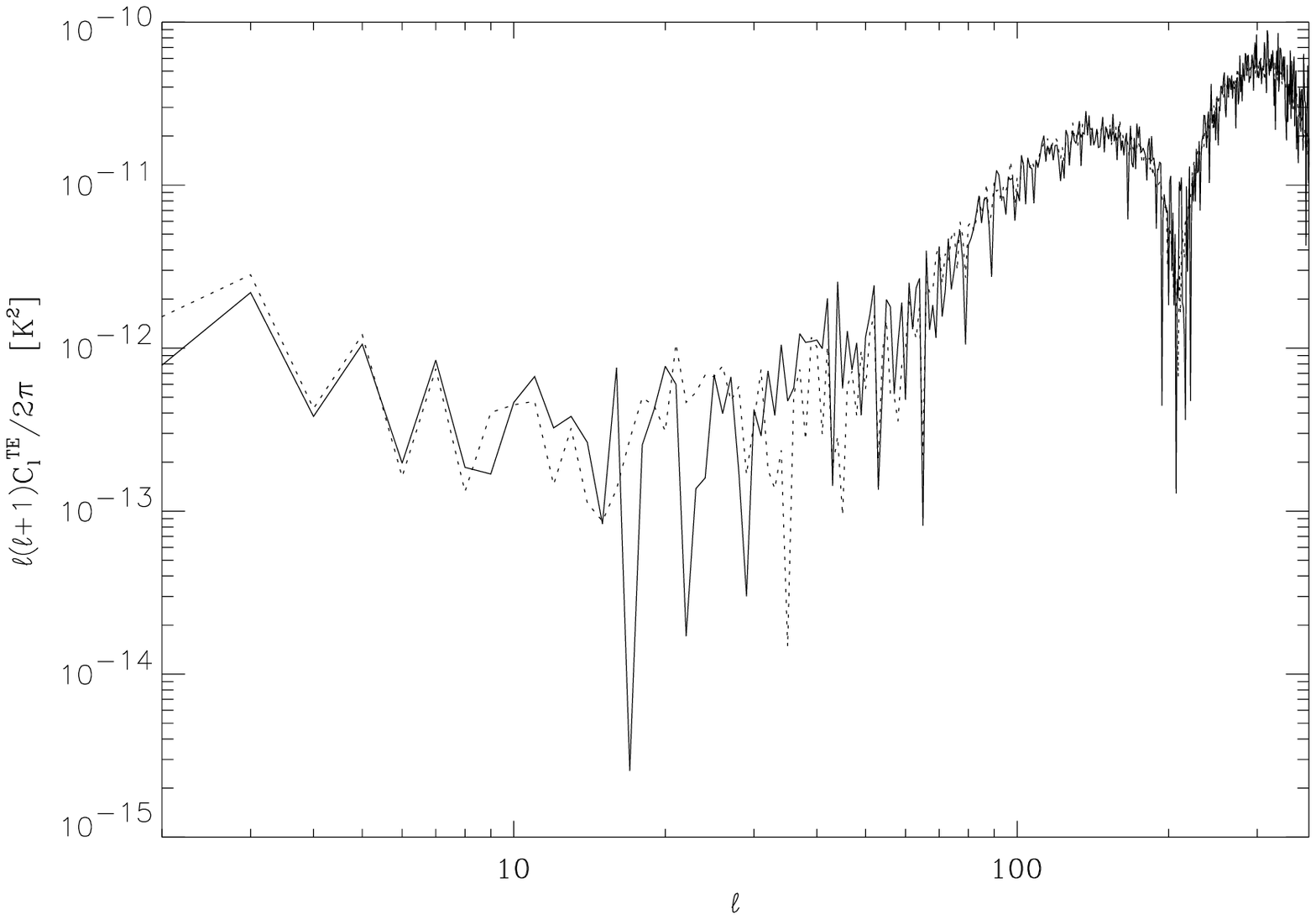,height=2.5in,width=3.in,angle=0}
\caption{$C_{\ell}^{TE}$ of original (dotted) and recovered
(solid) CMB emission obtained with {\ica} applied to simulated
{\sc Planck} maps, by considering the $S_{G}$ (left) and the
$S_{B}$ (right) foreground model, respectively. Results are shown
at 100 GHz.} \label{TElowl}
\end{center}
\end{figure*}

Both in the $S_{G}$ and in the $S_{B}$ case the synchrotron
contamination is almost absent in the acoustic oscillation region
of the spectrum, as it is evident again from Fig. \ref{TESYNCMB};
neglecting it we get the results shown in Fig. \ref{TEhighl}. The
combination of {\sc Planck} angular resolution and sensitivity
allows the recovery of the $TE$ power spectrum up to $\ell\simeq
1200$, corresponding roughly to the seventh CMB acoustic
oscillation.
\begin{figure*}
\begin{center}
\epsfig{file=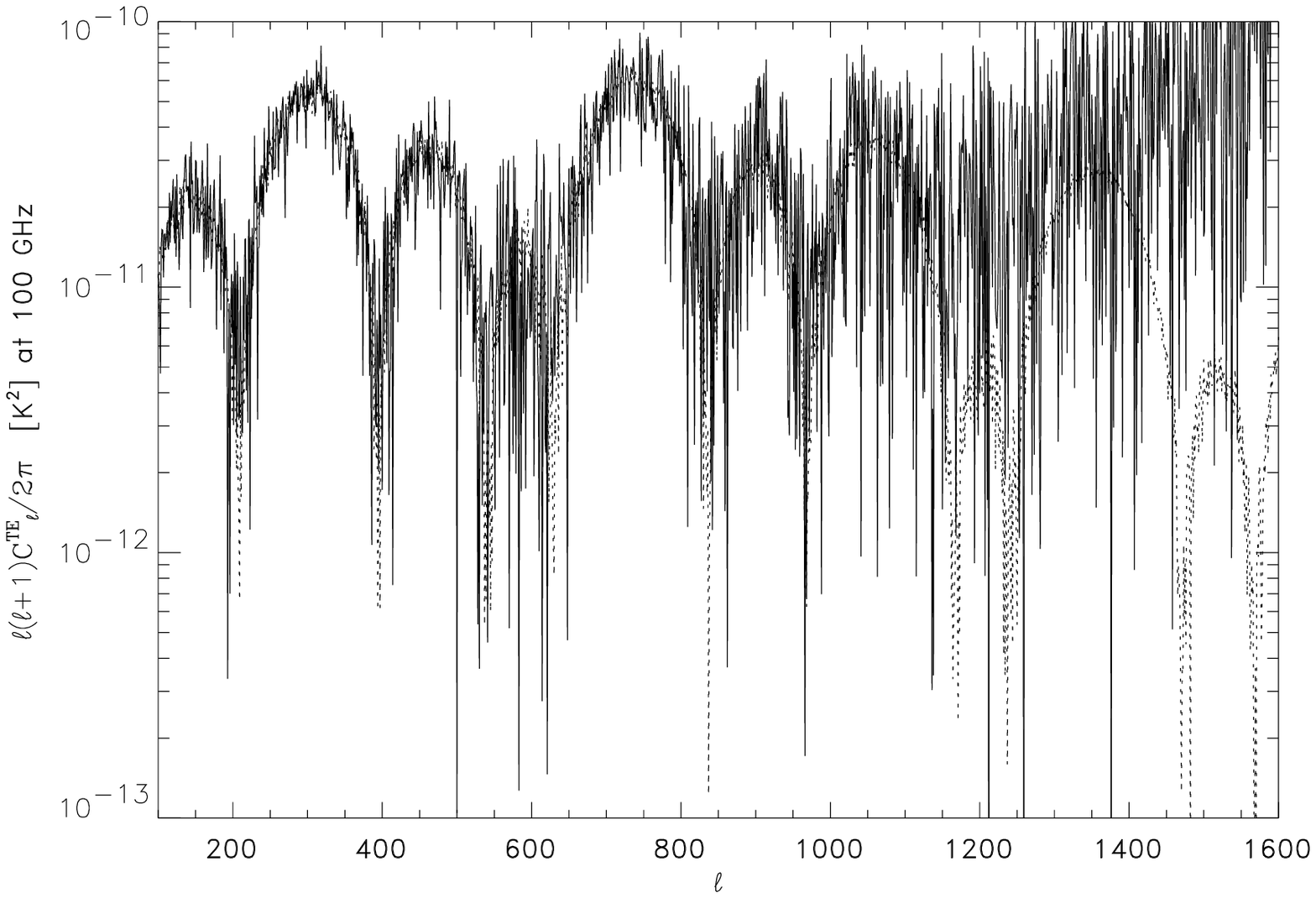,height=2.5in,width=3.in,angle=0}
\hskip .2in
\epsfig{file=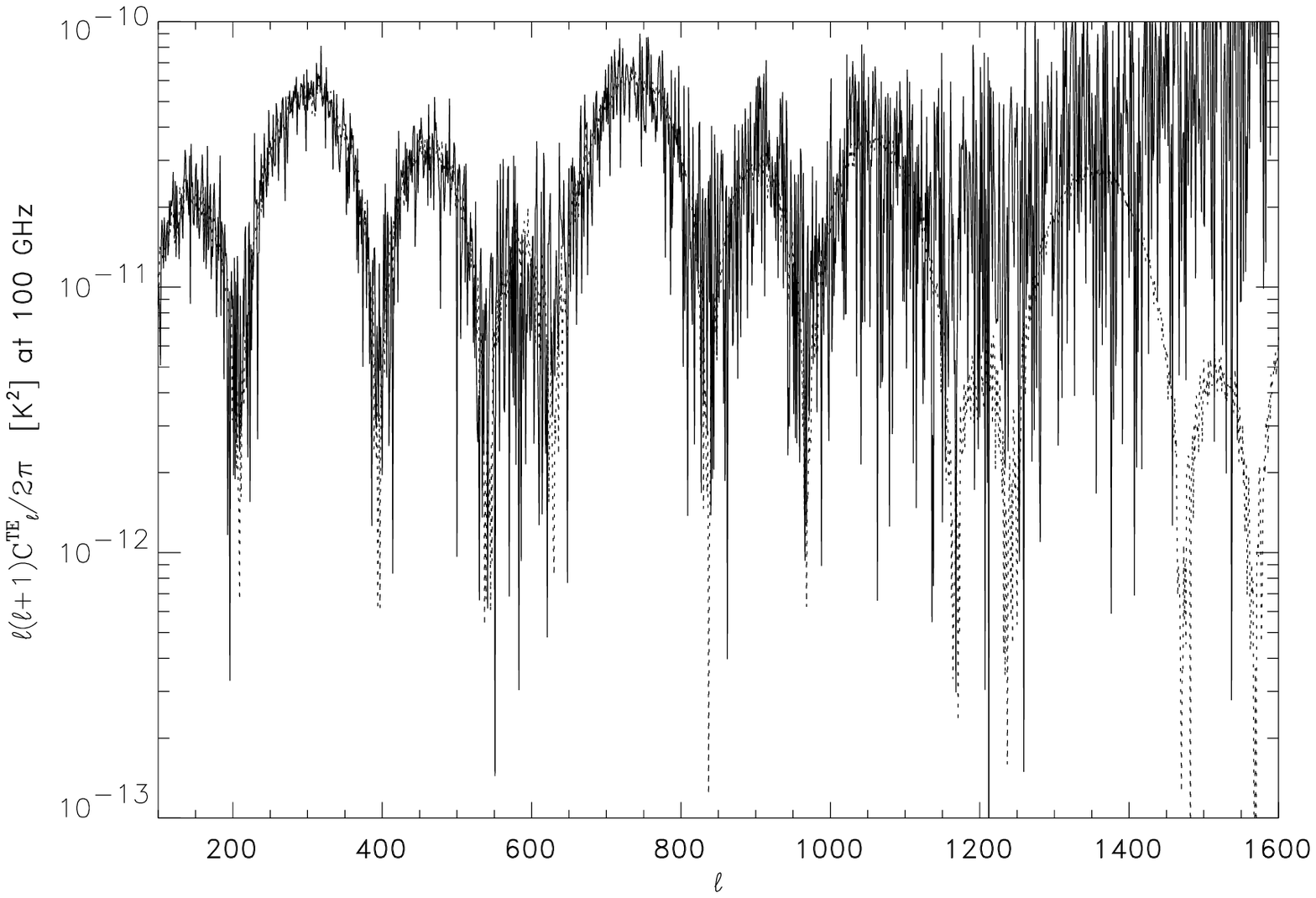,height=2.5in,width=3.in,angle=0}
\caption{$C_{\ell}^{TE}$ of original (dotted) and recovered
(solid) CMB emission adopting the $S_{G}$ (left) and the $S_{B}$
(right) synchrotron model, considering the {\sc Planck}
performances at 100 GHz.} \label{TEhighl}
\end{center}
\end{figure*}

\section{Concluding remarks}
\label{concluding}

Forthcoming experiments are expected to measure CMB
polarization\footnote{see lambda.gsfc.nasa.gov/ for a collection
of presently operating and future CMB experiments}. The first
detections have been obtained on pure polarization (Kovac et al.
2002), as well on its correlation with total intensity CMB
anisotropies, by the Wilkinson Microwave Anisotropy Probe (WMAP)
satellite\footnote{map.gsfc.nasa.gov/}.

The foreground contamination is mildly known for total intensity
measurements, and poorly known for polarization (see De Zotti 2002
and references therein). It is therefore crucial to develop data
analysis tools able to clean the polarized CMB signal from
foreground emission by exploiting the minimum number of a priori
assumptions. In this work, we implemented the Fast Independent
Component Analysis technique in an astrophysical context ({\ica}, 
see Amari, Chichocki 1998,
Hyv\"{a}rinen 1999, Baccigalupi et al. 2000, Maino et al. 2002) for 
blind component separation to deal with astrophysical polarized 
radiation.

In our scheme, component separation is performed {both on the
Stokes parameters $Q$ and $U$ maps independently and by joining
them in a single dataset}. $E$ and $B$ modes, coding CMB physical
content in the most suitable way (see Zaldarriaga, Seljak 1997,
Kamionkowski, Kosowsky, Stebbins 1997), are then built out of the
separation outputs. We described how to estimate the noise on
{\ica} outputs, on $Q$ and $U$ as well as on $E$ and $B$.

We tested this strategy on simulated polarization microwave all
sky maps containing a mixture of CMB and Galactic synchrotron. CMB
is modelled close to the current best fit (Spergel et al. 2003),
with a component of cosmological gravitational waves at the $30\%$
level with respect to density perturbations. We also included
re-ionization, although with an optical depth lower than indicated
by the WMAP results (Bennett et al. 2003a) since they came while
this work was being completed, but consistent with the
Gunn-Peterson measurements by Becker et al. (2001). Galactic
synchrotron was modelled with the two existing templates by
Giardino et al. (2002) and Baccigalupi et al. (2001). These models
yield approximately equal power on angular scales above the
degree, dominating over the expected CMB power. On sub-degree
angular scales, the Giardino et al. (2002) model predicts an
higher power, but still subdominant compared to the CMB $E$ mode
acoustic oscillations. Note that at microwave frequencies, the
fluctuations at high multipoles ($\ell \gsim 1000$), corresponding
to a few arcmin angular scales, are likely dominated by compact or
flat spectrum radio sources (Baccigalupi et al. 2002b, Mesa et al.
2002). Their signal is included in the maps used to estimate the
synchrotron power spectrum.

We studied in detail the limiting performance in the noiseless
case, as well as the degradation induced by a Gaussian, uniformly
distributed noise, by considering two frequency combinations: 30
$+$ 44 GHz and 70 $+$ 100 GHz. In the noiseless case, the
algorithm is able to recover CMB $E$ and $B$ modes on all the
relevant scales. In particular, this result is stable against the
space variations of the synchrotron spectral index indicated by
the existing data. In this case, {\ica} is able to converge to an
average synchrotron component, characterized by a ``mean" spectral
index across the sky, and to remove it efficiently from the map.
The output CMB map, also containing residual synchrotron due to
its space varying spectral index, is mostly good as far as the
frequencies considered are those where the synchrotron
contamination is weaker.

By switching on the noise we found that separation, at least for
what concerns the CMB $E$ mode, is still satisfactory for noise
exceeding the CMB but not the foreground emission. The reason is
that in these conditions the algorithm is still able to catch and
remove the synchrotron component efficiently. {We implemented
a Monte Carlo chain varying the CMB and the noise realizations in
order to show that the performance quoted above is typical and
does not depend on the particular case studied. Moreover, we
studied how the foreground emission biases the recovered CMB map,
by computing maps of residuals, i.e. subtracting the true CMB map
out of the recovered one. In the noiseless case, the residual is
just a copy of the foreground emission, with amplitude decreased
proportionally to the accuracy of the separation matrix. In the
noisy case, for interesting noise amplitudes the residual maps are
dominated by the noise in the input data, linearly mixed with 
the separation matrix. The situation is obviously worse for the weaker 
CMB $B$ mode. 

We applied these tools making reference to the {\sc Planck}
polarization capabilities, in terms of frequencies, angular
resolution and noise, to provide a first example of how the {\ica}
technique could be relevant for high precision large polarization
data-sets. We addressed separately the analysis of the CMB $E$,
$B$, and $TE$ modes. {While this work was being completed, the
Low Frequency Instrument (LFI) lost its 100 GHz channel, having
polarization sensitivity. However polarimetry at this frequency
could be restored if the 100 GHz channel of the High Frequency
Instrument (HFI) is upgraded, as is presently under discussion.
Due to the scientific content of the CMB polarization signal, the
{\sc Planck} polarization sensitivity deserves a great attention.
Within our context here, it is our intention to support the
importance of having polarization capabilities in all the
cosmological channels of {\sc Planck}, and in particular at 100
GHz. Our results have been obtained under this assumption.}

To improve the signal statistics, we found convenient to consider
at least three frequency channels in the separation procedure,
including the ones where the CMB is strongest, 70 and 100 GHz,
plus one out of the two lower frequency channels, at 30 and 44
GHz. Since the latter have lower resolution we had to degrade the
higher frequency maps since the present {\ica} architecture cannot
deal with maps having different resolutions. CMB $E$ and $TE$
modes were accurately recovered for both the synchrotron models
considered. The $B$-mode power spectrum is recovered on very large
angular scales in the presence of a conspicuous re-ionization
bump. On smaller scales, where the $B$-mode power mainly comes
from cosmological gravitational waves, the recovery is only
marginal for a $30\%$ tensor to scalar perturbation ratio.

On the sub-degree angular scales the contamination from
synchrotron is almost irrelevant according to both models
(Giardino et al. 2002, Baccigalupi et al. 2001). Moreover, it is
expected that Galactic $E$ and $B$ modes have approximately the
same power (Zaldarriaga 2001), while for CMB the latter are
severely damped down since they are associated with vector and
tensor perturbations, vanishing on sub-horizon scales at
decoupling corresponding to a degree or less in the sky (see Hu et
al. 1999). This argument holds also if the $B$ mode power is
enhanced by weak lensing effects from matter structures along the
line of sight (see Hu 2002 and references therein). Therefore, on
these scales, we expect the $E$ power spectrum to be a sum of
Galactic and CMB contributions, while the $B$ power comes
essentially from foregrounds only. In these conditions, the CMB
$E$ power spectrum is recovered by simply subtracting the $B$
power spectrum.

We also estimated the $TE$ contamination from synchrotron to be
irrelevant for CMB, because of the strength of the CMB $TE$
component due to the intrinsic correlation between scalar and
quadrupole modes exciting $E$ polarization. By applying these
considerations on sub-degree angular scales, as well as the
results of the {\ica} procedure described above on larger scales,
we show how the {\sc Planck} instrument is capable of
recovering the CMB $E$ and $TE$ spectra on all scales down to the
instrumental resolution, corresponding to a few arc-minutes
scales. In terms of multipoles, the $E$ and $TE$ angular power
spectra are recovered up to $1000$ and $1200$, respectively.

Summarizing, we found that the {\ica} algorithm, when applied to a
{\sc Planck}-like experiment, could be able to substantially clean
the foreground contamination on the relevant multipoles,
corresponding to degree angular scales and above. Since the
foreground contamination on sub-degree angular scales is expected
to be subdominant, the CMB $TE$ and $E$ modes are recovered on all
scales extending from the whole sky to a few arc-minutes. In
particular, the {\ica} algorithm can clean the $B$-mode power
spectrum up to the peak due to primordial gravitational waves if
the cosmological tensor amplitude is at least $30\%$ of the scalar
one. In particular, we find that on large angular scales, of a
degree and more, foreground contamination is expected to be severe
and the known blind component separation techniques are able to
efficiently clean the map from such contamination, as it is
presently known or predicted.

Still, despite these good results, the main limitation of the
present approach is the neglect of any instrumental systematics.
While it is important to assess the performance of a given data
analysis tool in the presence of the nominal instrumental
features, as we do here, a crucial test is checking the stability
of such tool with respect to relaxation of the assumptions
regarding the most common sources of systematic errors, like beam
asymmetry, non-uniform and/or non-Gaussian noise distribution
etc., as well as the idealized behavior of the signals to recover.
In this work we had a good hint about the second aspect, since we
showed that {\ica} is stable against relaxation of the assumption,
common to all component separation algorithms developed so far,
about the separability between space and frequency dependence for
all the signal to recover. In a forthcoming work we will
investigate how ICA based algorithms for blind component
separation deal with maps affected by the most important
systematics errors.

\section*{Acknowledgements}

C. Baccigalupi warmly thanks R. Stompor for several useful 
discussions. We are also grateful to G. Giardino for
providing all sky maps of simulated synchrotron emission (Giardino
et al. 2002), which has been named $S_{G}$ model in this paper. 
The HEALPix sphere pixelization scheme, available at {\tt www.eso.org/healpix}, 
by A.J. Banday, M. Bartelmann, K.M. Gorski, F.K. Hansen, E.F. Hivon, 
and B.D. Wandelt, jas been extensively used.

\end{document}